\documentclass[useAMS,usenatbib]{mn2e}
\usepackage{graphicx}    
\usepackage{amsmath}
\DeclareGraphicsExtensions{.ps,.pdf,.png}
\usepackage{natbib}
\makeatletter
\def\fps@figure{htbp}
\makeatother

\begin{document}
\title[The galaxy SMF in the UDS and CANDELS]{Deconstructing the Galaxy Stellar Mass Function with UKIDSS and CANDELS: the Impact of Colour, Structure and Environment}
\author[Mortlock et al.]{Alice~Mortlock$^{1,2}$, Christopher. J. Conselice$^{1}$, William G. Hartley$^{1,3}$, Ken Duncan$^{1}$,
\newauthor Caterina Lani$^{1}$, Jamie R. Ownsworth$^{1}$, Omar Almaini$^{1}$, Arjen van der Wel$^{4}$, 
\newauthor Kuang-Han Huang$^{5}$ Matthew L. N. Ashby$^{6}$, S. P. Willner$^{6}$, Adriano Fontana$^{7}$, 
\newauthor Avishai Dekel$^{8}$, Anton M. Koekemoer$^{9}$, Harry C. Ferguson$^{9}$, Sandra M. Faber$^{10}$,
\newauthor  Norman A. Grogin$^{10}$, Dale D. Kocevski$^{11}$
\footnotemark[0]\\
$^{1}$University of Nottingham, School of Physics and Astronomy, Nottingham, NG7 2RD UK \\ 
$^{2}$SUPA\thanks{Scottish Universities Physics Alliance} Institute for Astronomy, University of Edinburgh, Royal Observatory, Edinburgh EH9 3HJ\\
$^{3}$ETH Z\"urich, Institut f\"ur Astronomie, Wolfgang-Pauli-Str. 27, 8093, Z\"urich, Schweiz\\ 
$^{4}$Max-Planck Institut f\"ur Astronomie, K\"onigstuhl 17, D-69117, Heidelberg, Germany \\ 
$^{5}$Department of Physics, University of California Davis, One Shields Avenue, Davis, CA 95616, USA\\
$^{6}$Harvard-Smithsonian Center for Astrophysics, 60 Garden Street, Cambridge, MA 02138, USA\\ 
$^{7}$INAF - Osservatorio Astronomico di Roma, via Frascati 33,00040 Monte Porzio Catone, Italy\\
$^{8}$Racah Institute of Physics, The Hebrew University, Jerusalem 91904 Israel \\
$^{9}$Space Telescope Science Institute, 3700 San Martin Drive, Baltimore, MD 21218, USA \\ 
$^{10}$UCO/Lick Observatory, Department of Astronomy and Astrophysics, University of California, Santa Cruz, CA 95064, USA \\ 
$^{11}$Department of Physics and Astronomy, University of Kentucky, Lexington, KY 40506, USA}
\date{Accepted 12th November 2014}
\pagerange{\pageref{firstpage}--\pageref{lastpage}} \pubyear{2014}
\maketitle

\label{firstpage}

\begin{abstract}
We combine photometry from the UDS, and CANDELS UDS and CANDELS GOODS-S surveys to construct the galaxy stellar mass function probing both the low and high mass end accurately in the redshift range $0.3<z<3$.  The advantages of using a homogeneous concatenation of these datasets include meaningful measures of environment in the UDS, due to its large area (0.88 deg$^2$), and the high resolution deep imaging in CANDELS (H$_{160} > 26.0$), affording us robust measures of structural parameters.  We construct stellar mass functions for the entire sample as parameterised by the Schechter function, and find that there is a decline in the values of $\phi$ and of $\alpha$ with higher redshifts, and a nearly constant M$^{*}$ up to $z\sim3$. We divide the galaxy stellar mass function by colour, structure, and environment and explore the links between environmental over-density,  morphology, and the quenching of star formation. We find that a double Schechter function describes galaxies with high S\'ersic index ($n>2.5$), similar to galaxies which are red or passive. The low-mass end of the $n>2.5$ stellar mass function is dominated by blue galaxies, whereas the high-mass end is dominated by red galaxies. This hints that possible links between morphological evolution and star formation quenching are only present in high-mass galaxies. This is turn suggests that there are strong mass dependent quenching mechanisms. In addition, we find that the number density of high mass systems is elevated in dense environments, suggesting that an environmental process is building up massive galaxies quicker in over densities than in lower densities.
\end{abstract}

\begin{keywords}
galaxies: evolution--galaxies: formation--galaxies: luminosity function, mass function--galaxies: structure
\end{keywords}

\newpage

\section{Introduction}
\label{sec:int}
The galaxy stellar mass function (galaxy SMF) is an important tool for exploring galaxy evolution and the growth of stellar mass over cosmic time. By investigating the galaxy SMF of different populations we can begin to understand the physical processes which govern stellar mass growth as well as the types of galaxies affected by these physical processes.  With the advent of surveys which cover large areas, e.g. the Cosmic Evolution Survey (COSMOS; \citealt{Scov07}), we are now able to study the galaxy SMF with respect to different galaxy properties.

Some of the first studies which investigated the galaxy SMF focused on the local Universe. Both \citet{Cole01} and \citet{Bell03} constructed the local galaxy SMF down to stellar masses of  $M_{*}>10^{9} M_{\odot}$ by converting the $K-$band luminosity function. More recent studies have investigated the local galaxy SMF using stellar masses obtained from galaxy photometry (\citealt{LiWh09} and \citealt{Bald12}). These have pushed further down in stellar mass and computed the number densities of galaxies down to $M_{*}=10^{8} M_{\odot}$ and a good consensus has been reached on the shape of the local galaxy SMF.

Thanks to advances in technology, it is now possible to obtain large samples of galaxies for which redshifts and stellar masses can be computed. This has lead to the exploration of the stellar mass function across large ranges in redshift and stellar mass. Many of the earlier studies from the last decade have constructed the galaxy SMF up to redshifts of $z=1-2$ (e.g. \citealt{Font04}; \citealt{Cons05}; \citealt{Borc06}; \citealt{Bund06}; \citealt{Fran06}; \citealt{Pann06}). Other studies have pushed beyond this to measure the number densities of galaxies from $z=0$ to redshifts of $z\sim4-5$ (e.g. \citealt{Dror05}; \citealt{Font06}; \citealt{Pere08}) and higher (e.g. \citealt{Yan06}; \citealt{Star09}; \citealt{Capu11}; \citealt{Gonz11}; \citealt{Lee12}; \citealt{Dunc14}). These studies have provided excellent constraints on the evolution of the galaxy stellar mass function, often with deep data covering tens to hundreds of arcminutes squared.

In the last five years deeper data sets have facilitated better constraints on the low mass slope of the galaxy SMF. Studies which probe stellar masses down to $M_{*}\sim10^{9} M_{\odot}$ at redshifts of $z\sim1$ have found evolution in the low mass slope of the galaxy SMF such that it becomes steeper at higher redshifts (\citealt{Marc09}). Furthermore, several studies which probe the lowest mass populations have found that these systems are abundant and hence the low mass slope is very steep even at redshifts of $z=1$ (e.g. \citealt{Kaji09}; \citealt{Mort11}; \citealt{Sant12}; \citealt{Tomc14}). Complementing these results are studies which use data from large area surveys which is vital for constraining the high mass end of the galaxy SMF (e.g. \citealt{Ilbe10};  \citealt{Pozz10}; \citealt{Davi13}; \citealt{Ilbe13}; \citealt{Mous13}; \citealt{Muzz13}; \citealt{Tomc14}). The high mass turnover appears to evolve very little with redshift.

Numerous investigations have elucidated the dependence of build-up of stellar mass on different parameters. The dependance of the galaxy SMF on a system's star formation properties (often determined by colour) can clearly tell us lots about how galaxies are growing. Early studies showed that the stellar mass distribution of star forming and non star forming objects, or blue and red objects, are very different. At redshift of $z<1$ red objects were found to dominate the high mass end of the galaxy SMF whereas blue systems dominate the low mass end (\citealt{Borc06}; \citealt{Bund06}; \citealt{Bell07}). The first studies which explored the low mass end of the red SMF showed that the number densities of red low mass systems evolve quickly, resulting in a bimodality at $z<1$ (\citealt{Dror09}) and higher (\citealt{Ilbe10}). More recent studies have also found a bimodal form to the galaxy SMF of red galaxies, as well as surprisingly constant evolution in the SMF of blue systems (\citealt{Mort11}; \citealt{Bald12}; \citealt{Ilbe13}; \citealt{Mous13}; \citealt{Muzz13}).

The environment in which the galaxy resides is also a key parameter which affects the build-up of galaxy stellar mass. A galaxy in a dense environment may interact with the surrounding galaxies and may build up stellar mass through mergers or be quenched via gas stripping or merger driven feedback. This in turn can affect galaxy properties such as morphology and structure (e.g., \citealt{Dres80}; \citealt{Bamf09}; \citealt{Malt10}; \citealt{Skib12}; \citealt{Lani13}) or star formation properties and colour (e.g., \citealt{Coop07}; \citealt{Vand08}; \citealt{Chut11}; \citealt{Grut11a}; \citealt{Grut11}). However, it is difficult to discern whether it is the environmental processes or internal processes (e.g. in-situ star formation or feedback) which are driving mass growth and the truncation of star formation. By constructing the galaxy SMF with respect to environment and star formation history, it is possible to explore how both internal and external processes are driving the growth of galaxies of different stellar masses.

Measuring the environment of a galaxy is, however, a non-trivial problem which requires either a large survey area or targeted observations of cluster, group and field galaxies. Therefore there has only been a handful of work date which examines the galaxy SMF in different environments e.g.\citealt{Balo01}; \citealt{Koda03}; \citealt{Bund06}; \citealt{Bolz10}; \citealt{Vulc11}. The main conclusions of these studies is that the number densities of  galaxies in the highest density regions are dominated by massive, early type galaxies. This result is another way of exploring the already well established morphology-density relation (\citealt{Dres80}; \citealt{Post84}; \citealt{Treu03}; \citealt{Post05}; \citealt{Nuij05}; \citealt{Hold07}).

\citet{Peng10} find that out to redshifts of $z\sim1$ observations can be explained by the effects of environment and stellar mass being separable, and that these two effects halt star formation in galaxies. The evolution of the shape of the galaxy SMF of star forming and passive galaxies in the SDSS supports their proposed forms of quenching. Furthermore, the evolution of the UltraVISTA galaxy SMF of blue and red galaxies can also be explained as a mixture of mass and environment quenching (\citealt{Ilbe13}). The UKIRT Infrared Deep Sky Survey (UKIDSS, \citealt{Lawr07}) Ultra Deep Survey (UDS) is the ideal data set to test the effect of environment on the galaxy SMF, as the large area covered ($\sim$1 deg$^2$) allows us to probe different galaxy environments.

The build up of stellar mass in a galaxy also links with galaxy structure. Recent advances in observations allow us to probe galaxy structure at high redshift with high resolution imaging. By investigating the galaxy SMF with respect to galaxy structure we can infer how the morphologies of galaxies change as a function of stellar mass and redshift. Furthermore, several studies have previously noted that there is strong correlation between the presence of a bulge and passivity, (e.g. \citealt{Bell12}; \citealt{Bruc12}; \citealt{Cheu12}; \citealt{Barr13}; \citealt{Fang13};  \citealt{Bruc14}; \citealt{Lee13}; \citealt{Lang14}; and \citealt{Will14}). Although there is a significant passive disk population (e.g. \citealt{Mclur13} and \citealt{Bruc14}) which questions whether or not the presence of a bulge causes passivity (or vice versa). We can therefore use the galaxy SMF divided by structural parameters, compared to that divided by colour,  to further examine the possible link between the two. 

Previously, studies have investigated the galaxy SMF using various selections which correlate with morphology using various classification methods. These include visual morphology (\citealt{Cons05}), automated methods (\citet{Fran06}; \citealt{Pann06}; \citealt{Pann09}), and spectral type (\citealt{Font04} and \citealt{Verg08}). In \citet{Pozz10} a comparison between various classifications is performed and they found that there is not always agreement between different method of dividing systems by morphology. Nevertheless these studies all find that at $z<1$ the massive end ($M_{*}>10^{11} M_{\odot}$) of the galaxy SMF is dominated by the red, dead, and `well fed' systems with high S\'ersic indices ($n>2.5$). The low mass systems are dominated by star-forming, blue, low S\'ersic index ($n<2.5$) systems. Furthermore, the morphological mix changes with redshift such that at higher redshift more and more of the `late-type' objects dominate the stellar mass budget. Evidence for morphological division is also present in the $K-$band luminosity function. \citet{Deve09} reported evidence for the local number densities of bright objects being dominated by ellipticals and S0 galaxies, as well as fainter objects being dominated by disk type galaxies. Further investigation is needed  to understand how the shape of the galaxy SMF is changing for systems with different structure at high redshift. The high resolution, near infrared imaging from the Cosmic Assembly Near-infrared Deep Extragalactic Legacy Survey (CANDELS; PIs Faber and Ferguson; \citealt{Grog11} and \citealt{Koek11}) provides excellent data to perform this study. 

In this work we use the combined power of the UDS, CANDELS UDS and CANDELS GOODS-S data sets to explore the total galaxy SMF over several orders of magnitude in stellar mass up to redshifts of $z\sim3$. In addition, we also explore the evolution of the galaxy SMF with respect to colour, environment, and structure, drawing on the strengths of our three different surveys to provide the most robust measure for each. This paper is set out as follows. Section \ref{sec:datach2} describes how we calculate redshifts, stellar masses, rest frame colours, environment and structural parameters. Section \ref{sec:totgsmf} describes the total galaxy SMF from a combination of the UDS, CANDELS UDS and CANDELS GOODS-S data. Section \ref{sec:justcol} describes the galaxy SMF divided into passive and star forming galaxies. Section \ref{sec:justenv} describes the galaxy SMF divided into high and low density in the UDS. Section \ref{sec:structmf} discusses the galaxy SMF of galaxies in the the two CANDELS fields divided into galaxies with high and low S\'ersic index and asymmetry.  In Section 4 we interpret our results, and Section 5 summarises our findings. Throughout we assume $\Omega_{M}=0.3$, $\Omega_{\Lambda}=0.7$ and $H_{0}=70$ km s$^{-1}$ Mpc$^{-1}$. AB magnitudes and a Chabrier initial mass function (IMF) are used unless otherwise indicated.

\section{Data and Surveys}
\label{sec:datach2}
For this work a sample of galaxies at redshifts $0.3<z<3.0$ from the UDS DR8 dataset, as well as the CANDELS-UDS and  CANDELS GOODS-S data sets is selected. Using the combination of these two datasets we robustly probe the evolution of both the most massive galaxies over a wide area using the UDS, and fainter galaxies over a smaller area using the deep CANDELS UDS and GOODS-S data. We can take advantage of the different strengths of the two data sets, such as the wide area of the UDS which allows us to measure galaxy environment, and the high resolution CANDELS imaging which allows us to explore galaxy structure. Hereafter, we will refer to the UDS, CANDELS UDS and CANDELS GOODS-S datasets as the UDS, C-UDS and GOODS-S respectively.

\subsection{Photometry}
\subsubsection{The UDS}
\label{sec:udsphotom}
The UDS DR8 data reaches 5$\sigma$, 2$^{"}$-aperture depths of J=24.9, H=24.2 and K=24.6 and covers a total of 0.88 deg$^2$.  This survey contains over 100,000 galaxies with reliable redshifts, stellar masses and rest-frame colours. The area of the UDS allows us to probe different galaxy environments. For further information on the UDS see Almaini et al. (in prep).

In this study, SED fitting was used to computed photometric redshifts, stellar masses and rest frame colours. To perform this analysis robustly it is vital to have a large multiwavelength data set. The UDS benefits from additional wavelength coverage from various other surveys: Optical $B,V,R,i,z-$band data from the Subaru/XMM-Newton Deep Survey (SXDS; \citealt{Furu08}) and $U-$band data from the CFHT (Foucaud et al. in prep), 3.6 and 4.5 $\mu$m (IRAC channels 1 and 2) data from the UDS Spitzer Legacy Program (SpUDS; PI:Dunlop). The 24$\mu m$ data from SpUDS (limited to 300 $\mu$Jy; 15$\sigma$) is also used in this work for test purposes but not included in the SED fitting (see Section \ref{sec:uvjselec}). For further discussion of the data used in the UDS see Almaini et al. (2014) and \citet{Hart13}.

\subsubsection{CANDELS UDS}
\label{sec:canphotom}
CANDELS is a Multi Cycle Treasury Programme which images objects with both the Wide Field Camera 3 (WFC3) and the Advanced Camera for Surveys (ACS). In total, CANDELS consists of 902 orbits of the HST and covers 800 arcminutes$^2$. This area is made up of five different fields: GOODS-N, GOODS-S, EGS, COSMOS and UDS. CANDELS is divided into two parts CANDELS/Deep which images GOODS-N and GOODS-S to a 5$\sigma$ point source depth of H=27.7 mag, and CANDELS/Wide which images all fields to a 5$\sigma$ depth of H=26.3 in a 1$^{"}$-aperture. 

For this work, we use the CANDELS data that covers a part of the UDS field and the GOODS-S field, both of which have a pixel scale of $0.06$ arcseconds/pixel. The areas of the C-UDS and GOODS-S are approximately 0.06 and 0.05 deg$^2$  . For further details on the CANDELS data see \citet{Grog11} and \citet{Koek11}. This study utilises the UDS and GOODS-S photometry presented in \citet{Gala13} and \citet{Guo13} respectively.

As C-UDS is a subset of the UDS field it also benefits from the same wealth of ancillary data as the UDS dataset (Section \ref{sec:udsphotom}). However, in addition to the $U-$band CFHT data, $B,V,R,i,z-$band SXDS data and $J$, $H$ and $K$-band data from UKIDSS, the C-UDS data set also includes additional photometry. This includes F606W and F814W data from the Advanced Camera for Surveys (ACS), $H_{160}$ and $J_{125}$-band HST WFC3 data from CANDELS, $Y$ and $Ks$-bands taken as part of the HAWK-I UDS and GOODS-S survey (HUGS; VLT large program ID 186.A-0898, PI: A. Fontana; Fontana et al. in press). For the C-UDS, the 3.6 and 4.5$\mu m$ data is taken as part of the Spitzer Extended Deep Survey (SEDS; PI: G. Fazio,\citealt{Ashb13}). SEDS is deeper than SpUDS, which is used for in the UDS data set, but is only available over a 0.17 deg$^{2}$ region. Therefore, SEDS is a more appropriate choice for the much smaller C-UDS area. For a detailed discussion of the C-UDS field and the C-UDS photometry used in this work see \citet{Gala13}. 

\subsubsection{CANDELS GOODS-S}
\label{sec:gdsphotom}
The GOODS-S photometry consists of 17 bands from both ground and space based telescopes. This includes two $U-$bands, one taken with the MOSAIC II imager on the CTIO Blanco telescope and one with the VIMOS instrument on the VLT. There is also ACS F435W, F606W, F775W, F814W, F850LP data, as well as the WFC3 F098W, F105W, $J_{125}$ and $H_{160}-$band data. There are two sets of $K-$band photometry available in the GOODS-S. The first was taken using the ISAAC instrument on the VLT and the second was taken as part of the HUGS program using HAWK-I on the VLT. Finally, the photometry catalogue includes all four IRAC bands from SEDS and incorporates the pre- existing cryogenic observations from the GOODS Spitzer Legacy project (PI: M. Dickinson). For full detail of the GOODS-S data see \citet{Guo13} and references therein.

\subsubsection{Homogenisation}
It is important to consider how the data sets used in this study are homogenised. Correct homogenisation is crucial when combining data sets to ensure that total magnitudes and colours are in agreement between different data sets.

In the two CANDELS fields the software package \textsc{tfit} (\citealt{Laid07}) is used to derive all the photometry for non-HST data products. This package uses prior information on the position and morphology of any given object in a high resolution image to obtain the flux of that object in a lower resolution image. Part of the \textsc{tfit} process takes into account varying PSFs and hence one of \textsc{tfits} advantages is that it automatically provides correct colours. In the UDS corrections were applied to the two IRAC channels as well as the $U-$band data to obtain the correct colours. None of the other bands were corrected as the PSF sizes in the remaining bands are very similar (0.77-0.84 arcsec; \citealt{Simp12}). In short, the corrections were computed by using the $K-$band (for IRAC) or $B-$band (for the $U-$band) to estimate the flux lost due to the larger PSFs. For the $U-$band the corrections were only typically of 1\%, whereas for the two IRAC bands the correction could be between 26 and 30\%.

In all of the fields used in this work total fluxes are estimated using the \textsc{sextractor} parameter FLUX\_AUTO. For the two CANDELS fields all HST bands are converted to total fluxes using the ratio of FLUX\_AUTO$/$FLUX\_ISO. This was done prior to the \textsc{tfit} process, and hence the resulting photometry is total. In the UDS the conversion to total is applied to the stellar masses and rest frame magnitudes after photometry fitting using the ratio of FLUX\_AUTO$/$FLUX\_APER. Good agreement is found between the total $K-$band fluxes measured in the UDS and C-UDS. The median difference is 0.07 $\mu$Jy which corresponds to a median stellar mass difference of 0.03 dex.

\subsection{Photometric redshifts}
\label{sec:photoz}
\begin{figure*}
\centering
\includegraphics[trim = 22mm 0mm 0mm 6mm, clip,scale=0.65]{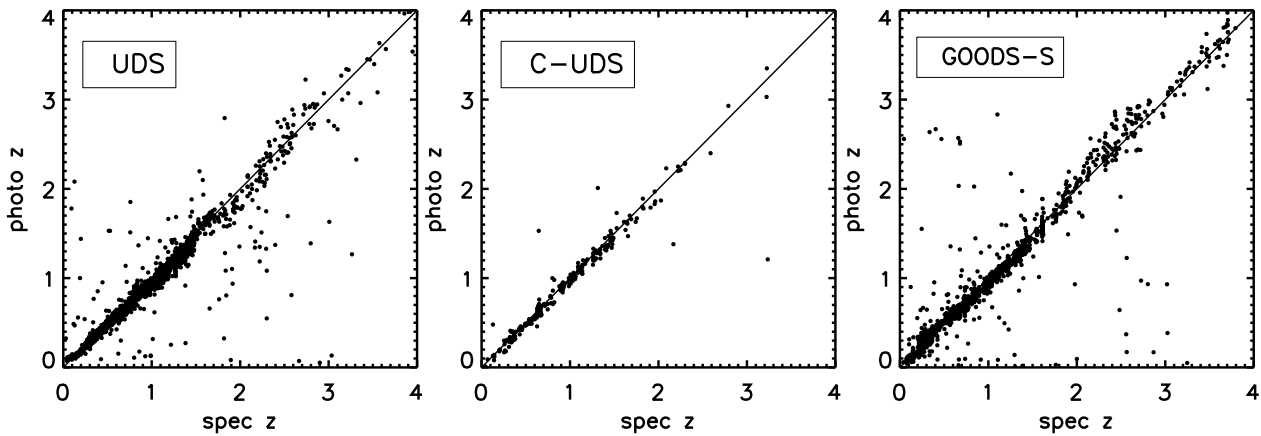}
\caption[Comparisons of spectroscopic and photometric redshifts.]{Left panel: Spectroscopic redshifts versus photometric redshifts for 2146 galaxies in the UDS with highest-confidence spec-z after excluding AGN. Middle panel: Spectroscopic redshifts versus photometric redshifts for 285 galaxies in the C-UDS which have spectroscopic redshifts from UDSz. Right panel: Spectroscopic redshifts versus photometric redshifts for 1840 galaxies in the GOODS-S which have spectroscopic redshifts as described in \citet{Dahl13}.}
\label{speczphotoz}
\end{figure*}
\subsubsection{The UDS photometric redshifts}
Photometric redshifts for the UDS sample were computed by \citet{Hart13}. They were determined via fitting template spectral energy distributions (SEDs) to the photometric data points described in Section \ref{sec:udsphotom} using the EAZY code (\citealt{Bram08}). The photometry was fit to the linear combinations of the six default EAZY templates, and an additional template which is the bluest EAZY template with a small amount of SMC-like extinction added ($A_{v}=0.1$). The redshifts were from maximum likelihood analysis. For full details of the fitting procedure and resulting photometric redshifts see \citet{Hart13}.

A comparison of the photometric redshifts used in this work to spectroscopic redshifts that are available in the UDS was carried out in \citet{Hart13}. Spectroscopic redshifts are from the UDSz, an ESO large spectroscopic survey (ID:180.A-0776) within the UDS. The spectroscopic redshifts within UDSz are for K-selected galaxies ($K_{AB} < $23.0) which were chosen to be at redshifts of $z>1$ and also include a low-redshift control sample. The UDSz spectroscopyc redshifts are combined with spectroscopic redshifts from the literature (see \citealt{Simp12} and references therein). From these two sources, and after exclusion of AGN, \citet{Hart13} used 2146 spectroscopic redshifts for comparison to photometric redshifts. Excluding catastrophic outliers ($\Delta z/(1+z)>0.15$, $\sim$3\%), the dispersion of $z_{photo} - z_{spec}$ is $\Delta z/(1+z)=0.03$ for this data. The left hand panel of Figure \ref{speczphotoz} shows the spectroscopic redshifts versus the photometric redshifts for the 2146 galaxies used to calculate the quality of our photometric redshifts. Objects with spectroscopic redshifts are the brightest subset of galaxies and for fainter systems the dispersion will be larger. This is discussed further in Section \ref{sec:masses} where a comparison between the UDS and C-UDS is made.

\subsubsection{The CANDELS UDS and GOODS-S photometric redshifts}
The C-UDS photometric redshifts were calculated using the same method as for the UDS sample described above. For the C-UDS, the SED templates are fit to the photometry described in \ref{sec:canphotom} and the best fit redshift is used. The middle panel of Figure \ref{speczphotoz} shows the spectroscopic redshifts versus the photometric redshifts for the 285 CANDELS galaxies which have spectroscopic redshifts (see \citealt{Gala13} for details of these spectroscopic redshifts). The dispersion of $z_{photo} - z_{spec}$ is $\Delta z/(1+z)=0.026$ for our C-UDS photometric redshifts (after removal of the 2\% of catastrophic outliers), although it is noted that only have a small sample of spectroscopic redshifts to compare to within the C-UDS region.

The photometry used to compute the GOODS-S photometric redshifts are described in \ref{sec:gdsphotom} and the method, which uses the probability weighted redshift, is described in detailed in \citet{Dunc14}. Figure \ref{speczphotoz} shows a spectroscopic versus photometric redshift comparison for 1840 of the GOODS-S galaxies which have spectroscopic redshifts. These spectroscopic redshifts are from a compilation of data from GOODS-S which is described in full in \citet{Dahl13} and references therein. We find a dispersion of $z_{photo} - z_{spec}$ is $\Delta z/(1+z)=0.036$ when catastrophic outliers (4\%) are removed as before. 

\citet{Dahl13} looked in detail at the photometric redshift analysis of the CANDELS fields and discussed the merits of combining different photometric redshifts for individual objects from multiple techniques to produce the best final redshift. For both the C-UDS and GOODS-S, the scatter and catastrophic outlier fractions are comparable to those produced with the optimised CANDELS photometric redshifts. Therefore we are confident that our photometric redshifts are of the highest possible quality.

\subsection{Stellar Masses and Rest Frame Magnitude}
\label{sec:masses}
\begin{figure*}
\centering
\includegraphics[trim = 22mm 0mm 0mm 6mm, clip,scale=0.65]{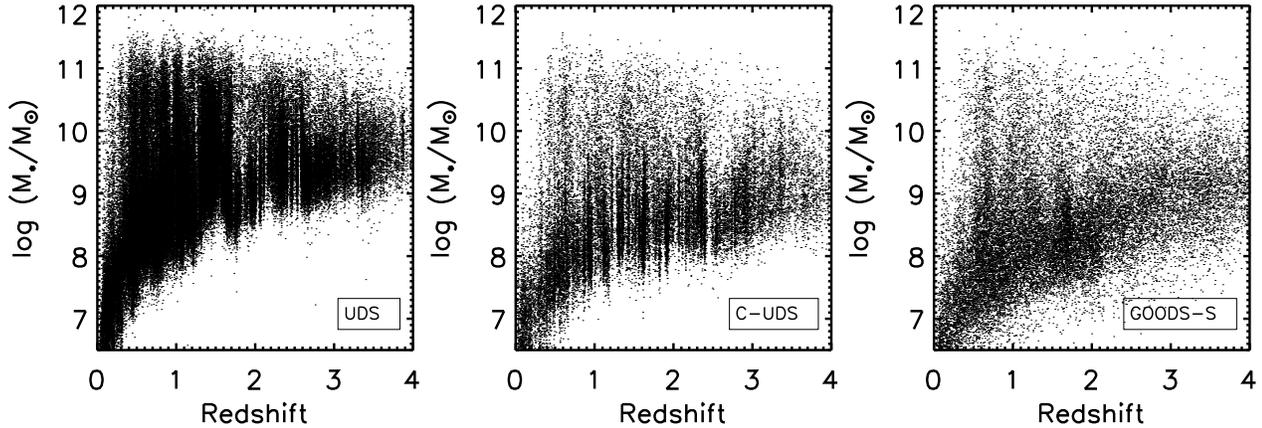}
\caption[Stellar mass as a function of redshift.]{Left panel: The stellar masses as a function of redshift for all galaxies in the UDS. Middle panel: The stellar masses as a function of redshift for all galaxies in the C-UDS. Right panel: The stellar mass as a function of redshift for all galaxies in GOODS-S.}
\label{redshiftmass}
\end{figure*}

\begin{figure*}
\centering
\includegraphics[trim = 0mm 0mm 0mm 0mm, clip,scale=0.65]{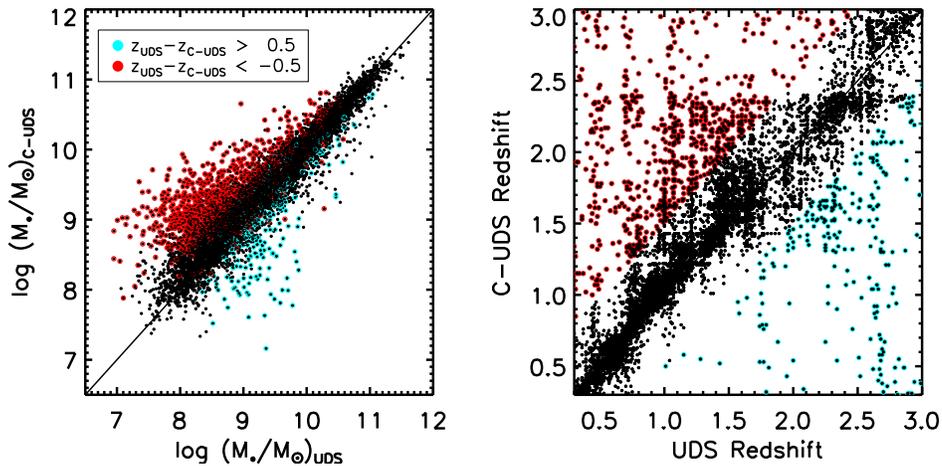}
\caption{Left panel: the stellar masses from the UDS field vs. the stellar masses of the C-UDS field (black points). Right panel: The photometric redshifts for the UDS field vs. the C-UDS field. Over plotted in red and blue are galaxies with large differences in photometric redshifts measured from the two different sets of photometry ($\sim11\%$ of galaxies with $0.3<z<3$).}
\label{redshiftcomp}

\end{figure*}
The method used to compute the stellar masses and rest frame magnitudes used in this work is described in detail in \citet{Mort13}; \citet{Hart13} and \citet{Lani13}. In brief, a large grid of synthetic SEDs was constructed from the stellar population models of Bruzual \& Charlot (2003, hereafter BC03), assuming a Chabrier IMF (\citealt{Chab03}). The reddening law of \citet{Calz00} was used to include dust in the templates. The star formation history was characterised by an exponentially declining model with various ages, metallicities and dust extinctions.
These models were parametrised by an age of the onset of star formation, and by an e-folding time such that
\begin{equation}
SFR(t) \sim SFR_{0}\times e ^{-\frac{t}{\tau}}.
\label{eq:SFRexpc2c2}
\end{equation}
\noindent where the values of $\tau$ ranged between 0.01 and 10.0 Gyr, while the age of the onset of SF ranged from 0.001 to 13.7 Gyr. The metalicity ranged from 0.0001 to 0.05 (BC03), and the dust content was parametrised by $\tau_{v}$, the effective $V-$band optical depth for which we used values $\tau_{v}$ = 0, 0.4, 0.8, 1.0, 1.33, 1.66, 2, 2.5, 5.0. Other star formation histories are not investigated in this work as studies have shown that stellar mass calculations are generally robust to changes in star formation history within our redshift range and for the stellar masses probed in this work (e.g., \citealt{Owns12}; \citealt{pfor12}; \citealt{Redd12}; \citealt{Ilbe13}).

Each template was scaled to the apparent magnitude of the detection band magnitude of the galaxy. For the UDS we used the $K-$band magnitude, and for the C-UDS and GOODS-S the $H_{160}-$band was used. The $\chi^{2}$ was then computed for each template and from the best fit template best fit stellar masses and best fit rest frame magnitudes were obtained. A modal mass value was also calculated by binning the stellar masses of the ten percent of templates with the lowest $\chi^{2}$ in bins of 0.05 dex. The mode of the stellar mass corresponds to the stellar mass bin with the largest number of templates. In this analysis we use the mode of the stellar masses and the best fit rest frame magnitudes as discussed in \citet{Mort13}.

Uncertainties on the stellar masses and rest frame colours were computed from a Monte Carlo analysis. For each galaxy, simulated photometry was created by randomly choosing a point on a Gaussian curve whose standard deviation is the measured uncertainty on the photometry. This was done for each input band, and then the stellar mass and rest frame colours from the simulated photometry were computed. This was then repeated 500 times and the uncertainty was taken as the standard deviation of the 500 Monte Carlo realisations. Furthermore, uncertainties due to the photometric redshift measurements were computed using the same method. The uncertainty from the photometry and the uncertainty from the photometric redshift were summed in quadrature and this was taken as the final uncertainty on the stellar masses. The stellar masses as a function of redshift are shown in Figure \ref{redshiftmass} for all galaxies in the UDS (left panel), the C-UDS (middle panel) and GOODS-S (right panel). The cut off in stellar mass in this figure highlights the difference in survey depths. 

Several studies have shown that the high redshift systems can exhibit SEDs which are better fit with the inclusion of nebular emission lines, and that these fits which include nebular emission result in lower stellar masses and ages (\citealt{Zack08}; \citealt{Scha09}; \citealt{Rait10}; \citealt{Scha10}; \citealt{Mclu11}; \citealt{Ono12}, \citealt{Curt13}; \citealt{Dunc14}). These studies focus on redshift ranges much higher than what is explored in this work, however it important to investigate any potential impacts this may have on our results. As the impact from nebular emission is expected to be on objects with lower stellar masses and younger ages we test the GOODS-S sample as this provides the strongest constraint on the the low mass end of the galaxy SMF.

We recompute the stellar masses of the GOODS-S sample including nebular emission in the SED templates (see \citealt{Dunc14} for discussion of the prescription for including nebular emission). For any given stellar mass bin, at any given redshift, the mean difference between the stellar mass with and with out nebular emission is always a factor of 10 to 100 smaller than the mean uncertainty on the stellar masses in that mass and redshift bin. The only exceptions to this are in the redshift range $2.0<z<2.5$ where we find mean differences of $\sim$0.2 (i.e. of the order of the uncertainties on the stellar masses) in the stellar mass range $M_{*}=10^{7.5} - 10^{8.5} M_{\odot}$. However, due to our completeness limits, galaxies of these stellar masses are not included in our SMFs in this redshift range. It is also possible that some low mass, low metallicity objects have such large emission lines that their photometry is boosted and they are therefore fit as high mass systems. We compute the difference between the stellar masses with and without nebular emission and find that no galaxy changes mass by more than $\sim$1.5 dex. Furthermore, the objects with this large difference are found only at the low mass end of the galaxy SMF ($M_{*}\sim 10^{8} M_{\odot}$). These objects will not be contaminating the high mass end of the galaxy SMF where the number density can be changed greatly by the inclusion/loss of just one object. Therefore we conclude that nebular emission lines do not impact our findings. These findings are in agreement with Santini et al. (submitted) who performed a comparison of galaxy properties derived from SED fits to a variety of model, with varying parameters, for the same GOODS-S galaxies which are used in this comparison. In terms of differences due to nebular emission, they find that only a sub-sample of young galaxies are affected and that the impact is harshest in the redshift range $2.2<z<2.4$ (as well as in several outside the range of this work).

\subsubsection{Redshift and Stellar Mass Differences}
\label{sec:zandmdiff}
The left hand panel of Figure \ref{redshiftcomp} shows the comparison between the UDS stellar masses and the C-UDS stellar masses for our sample where the two field overlap. Over plotted in blue are galaxies for which $z_{(UDS)}-z_{(C-UDS)} >  0.5$, where $z_{(UDS)}-z_{(C-UDS)}$ is the difference in the redshift computed from the UDS photometry and the redshift computed from the C-UDS photometry. This is found to be $\sim3\%$ of the galaxies in the redshift range $0.3<z<3.0$. The red symbols are galaxies for which $z_{(UDS)}-z_{(C-UDS)} < -0.5$ ($\sim8\%$ of the full data set in the redshift range $0.3<z<3.0$) and hence any major differences in stellar masses are a result of disagreements in the photometric redshifts. This can be seen in the right hand panel of Figure \ref{redshiftcomp}, which shows the comparison between the UDS and C-UDS photometric redshifts. 

Although superficially the spread appears large, only 11\% of the sample is discrepant by more than 0.5 in redshift and these are the galaxies which are close to the detection limit in the UDS. Therefore, a) a change in the photometric redshifts of the faintest objects is not necessarily a surprise due to the improved photometry for the deeper data and b) there are many galaxies at the low mass end and hence a small change in number will have less impact. Furthermore, the different depths of mid infra-red data used for the two data sets (see Section \ref{sec:canphotom}) will impact our photometric quantities. Despite the apparently large scatter it is noted that the median offset in each redshift bin is acceptable. The offset ranged from -0.03 in the lowest redshift bin to 0.18 in the highest redshift bin. The median offset in stellar mass in the lowest redshift bin is -0.02, which increases to 0.1 in the highest redshift bin. Hence, on average, we are finding good agreement between these two surveys.

It is important to consider how these problem galaxies (galaxies for which the difference in UDS and C-UDS redshift is more than 0.5) impact the galaxy SMF as a function of stellar mass. As previously mentioned many of the problem galaxies are near the magnitude limit of the UDS and include shallower IRAC data in the UDS. As a test the assumption is made that the C-UDS redshifts and masses are closer to the true redshifts and stellar masses for the specific problem galaxies. We stress that there is no reason to assume this is the case for brighter galaxies and also that there are clearly scenarios in which the properties derived from the C-UDS are not the best answer (e.g. some image problem resulting in bad photometry).

In each redshift bin, as defined by the C-UDS redshifts, there are a given number of galaxies which have could have problem redshifts (difference in UDS and C-UDS redshift is greater than 0.5) and hence masses in the UDS. However, as many of these galaxies are faint they are below the UDS stellar mass limit and hence do not contribute to the SMF. Furthermore many of these objects are below the UDS stellar mass limit according to both their UDS and C-UDS stellar masses, further evidence that they should not contribute to the SMF. Taking this into account we find that in our six redshift bins (z$\sim0.4,0.75,1.25,1.75,2.25,2.75$) the percentages of problem galaxies are 0$.$27\%, 0$.$31\%, 4$.$13\%, 10$.$77\%, 17$.$15\% and 5$.$0\% respectively. For most of the redshift bins this percentage is encouragingly small. Further to this, when comparing the UDS and C-UDS properties of these systems at redshifts of $z>1$ there is no trend such that a specific redshift or mass always moves to another redshift or mass bin. At redshift of $z<1$ there is a tendency for the redshifts and stellar masses of the problem objects to be higher in the UDS, however the contamination percentage is so low that this is unlikely to affect the shape of the galaxy SMF.

Finally, there are two specific problem regions to consider further. The first is the redshift range $1.5<z<2.5$ where the contamination percentage is high. The second is the high mass end of the galaxy SMF where the introduction or removal of a small number of objects can make a large difference. In the redshift range $1.5<z<2.5$ it is a worry that the high percentage of problem objects may affect the shape of the galaxy SMF. However, if this was the case there would be a difference in the UDS and C-UDS number densities which would be obvious by inspection. No such difference is found except for a small possible discrepancy in normalisation between the two fields. This may be explained by galaxies which have been scattered out of the UDS number densities into other redshift and stellar mass bins.

At the high mass end of the galaxy SMF there is only 1 galaxy within the sample of problem galaxies which is massive ($M_{*}>10^{11} M_{\odot}$) in the UDS (in the redshift bin $2.5<z<2.0$) but not the C-UDS. As a simple test as to how much this potentially changes the high mass end, it is assumed that there are 11 objects in this bin which are contamination (as the UDS is $\sim$11 times the area of the C-UDS). If this number is subtracted from the actual number of objects in that bin we find that the resulting number density is still within error of the previous result. We therefore argue this has no significant effect on our results at the high mass end (for further discussion see Section \ref{sec:eddybias}). It is important to note that the fainter galaxies in the UDS, which are the drivers behind the large spread, will also have larger uncertainties on their photometric redshifts and stellar masses. As these uncertainties are accounted for in our analysis (see Section \ref{sec:totgsmf}), the difference in photometric redshifts and stellar masses will be represented in the uncertainties on any quantity computed in this work.

\subsubsection{$UVJ$ Selection}
\label{sec:uvjselec}
To divide our galaxies into red and blue (passive and star forming) populations we used rest-frame $UVJ$ colours as in \citet{Wuyt07}. \citet{Will09} applied this method to the UDS DR1 data release and found that galaxies can be divided robustly into red/blue populations using the $UVJ$ criteria up to redshifts of $z\sim2$. Furthermore, \citet{Hart13} showed that this can be extended to redshifts of $z>2$ using the greater number statistics of the UDS DR8. We therefore adopt the following criteria to divide our sample in to red and blue systems. A galaxy is considered red, or passive, if 

\begin{equation}
(U-V) > 1.3~~~and~~~(V-J) < 1.6
\label{eq:uvj1}
\end{equation} 
\noindent with the additional redshift dependent criteria that:

\begin{align*}
\label{UVJ}
(U-V)&> \left\{
\begin{array}{ll}
&0.88\times(V-J) + 0.69~~~~~~0.3<z<0.5 \nonumber \\
\\
&0.88\times(V-J) + 0.59~~~~~~0.5<z<1.0 \nonumber \\
\\
&0.88\times(V-J) + 0.49~~~~~~1.0<z<2.5  \\
\end{array}
\right.
\end{align*}
\\
\noindent where $U-V$ and $V-J$ are the rest frame colours computed from the magnitudes discussed in Section \ref{sec:masses}

The $UVJ$ criteria are often used to distinguish between passive and star forming objects. However, dusty, star forming galaxies will contaminate the passive region, and hence the selection is not perfect. \citet{Hart13} used an additional cut to remove star forming objects from the passive sample, although they found that this had very little impact on their results. Our sample was tested for this by removing all galaxies with a 24$\mu$m detection from the passive sample. In our redshift range, and for any given stellar mass bin, the change in number density is on average $\sim$2\% (maximum change to a single stellar mass bin is $\sim$15\%). It is noted that much of the 24$\mu$m data is at redshifts of $z<1$ but we can say that there is no evidence for contamination given the data available to us.

To fully understand the impact of contamination arising from the UVJ selection on our results we would ideally need to know how the contamination changes as a function of stellar mass and redshift. However, without a considerable amount of spectra this is difficult to test. Previous studies have looked at how star formation rates correlate with position on the UVJ diagram. \citet{Bram11} showed that the UVJ selection criteria can clearly separate objects which are star-forming or quiescent based on their UV$+$IR star formation rates up to redshifts of $z=3$ and down to stellar masses of $M_{*}\sim10^{10} M_{\odot}$. Furthermore, \citet{Wuyt07} showed that systems in the quiescent part of the UVJ diagram, with stellar masses down to $M_{*}\sim10^{9} M_{\odot}$, are consistent with being old passively evolving populations based on their properties from SED fitting.

There have been a handful of studies which spectroscopically demonstrate the reliability of the UVJ selection. For example, \citet{Will09} showed that a sample of 96 old, passively evolving galaxies, which are spectroscopically confirmed, lie on the correct place on the UVJ diagram. These objects have redshifts of $z<2.5$ and $K-$band magnitudes brighter than $K<$22.4. \citet{Whit13} showed that the UVJ selection is very effective at selecting spectroscopically confirmed passive galaxies from the 3D-HST grism survey out to redshifts of $z\sim2$ and most recently \citet{Bell14} have used spectroscopic properties to show that galaxies which lie in the quiescent region of the $UVJ$ diagram are genuinely quiescent (e.g. not contamination from dusty star forming galaxies). This sample of galaxies is in the redshift range $1<z<1.6$ and has stellar masses of $M_{*}>10^{10.7} M_{\odot}$.

It is important to note that the uncertainties on the number densities of red and blue objects include consideration of the scattering of objects around the UVJ diagram due to errors on the rest-frame magnitudes (e.g. due to uncertainties in the photometry). Therefore we believe the average contamination is represented in the results presented for the blue and red SMFs. However we can perform a similar test, as in Section \ref{sec:zandmdiff}, where we consider the comparison between the UDS and C-UDS $UVJ$ classifications. In this way we can explore how galaxies move around the $UVJ$ diagram due to differences in these two data sets. For blue (or star forming) systems in the C-UDS the percentage of objects which have a different $UVJ$ classification in the UDS is not a strong function of mass or redshift. At $M_{*}>10^{11} M_{\odot}$ the contamination percentage is zero at all redshifts. For stellar masses of $M_{*}<10^{11} M_{\odot}$ down to the stellar mass limit of the UDS the percentage is generally of the order of a few to ten percent at all redshifts. However, there are some stellar mass bins where it can be as high as 18\%.

For the passive or red galaxies the change in classification does seem to be a function stellar mass such that it gets progressively worse for intermediate stellar masses. Considering objects with $M_{*}>10^{11.5} M_{\odot}$ there is no shift, however when looking at objects with stellar masses of $10^{11}<M_{*}<10^{11.5} M_{\odot}$ the change ranges from as little as 5\% to as much as 25\% in the different redshift bins. At stellar masses of $10^{10}<M_{*}<10^{11} M_{\odot}$ the average change is $\sim$20\% (however it ranges from zero change to as large as $\sim$40\% in one stellar mass bin). For stellar mass bins of $M_{*}<10^{10} M_{\odot}$ the average shift is again $\sim$20\% however there is a smaller spread across the different redshift ranges (as low as zero and as high as $\sim$30\% in one stellar mass bin.) Although there does seem to be a general trend of the contamination getting worse at higher redshift the trend is not smooth.

Some caution needs to be taken when considering these results (as with the results in Section \ref{sec:zandmdiff}). It is not necessarily true that the C-UDS gives us the `correct' answer when compared with the UDS. There are other factors which need to be considered which could mean that for a given object the UDS photometry is more accurate (e.g. image quality at a given part of the detector or distance from the edge of the image). When considering galaxies which have a different UVJ classification in UDS to C-UDS at a given stellar mass, a given redshift, and a given colour small number statistics begin to impact these results. There are often only of the order of tens of objects which have a different UVJ classifications between the two data sets and hence if the assumption that C-UDS is `correct' is wrong for just a few then this can change the percentages reported here. It is worth noting that the similarity between the number densities for red and blue objects, in the UDS and the C-UDS, indicates that any potential contamination does not seem to have a large impact upon our results.

\subsection{Completeness}
\label{sec:compcalc}
\begin{figure*}
\centering
\includegraphics[trim = 22mm 0mm 0mm 0mm, clip,scale=0.65]{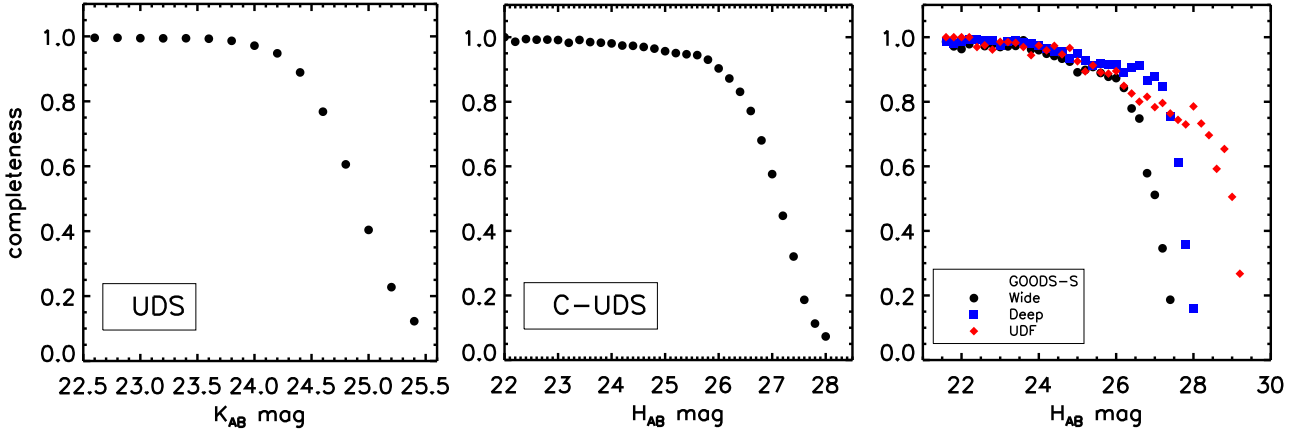}
\caption[The completeness curves as a function of magnitude for the UDS and C-UDS.]{The left, middle and right panels are the completeness curves as a function of magnitude for the UDS (Almaini et al. 2014), the C-UDS, and the three regions in the CANDELS GOODS South (Yan et al. 2013) respectively. In the right hand panel, the black symbols are for the wide region of GOODS South, the blue symbols are for the deep region and the ERS region, and the red symbols are for the HUDF.}
\label{compcurve}
\end{figure*}

As low stellar masses are probed at high redshift, number densities become incomplete as the faintest galaxies cannot be observed. Using the detection completeness as a function of magnitude (shown in Figure \ref{compcurve}) we corrected the number densities for the total galaxy SMFs.  Figure \ref{compcurve} shows the completeness for the three surveys used in this work. The horizontal axis of the left panel is $K-$band and the middle and right panel are $H-$band. For the galaxies used in this study the typical observed $H-K$ colours are $\sim0.5$. Here we describe briefly the methods for calculating the completeness functions.

\noindent {\bf UDS:} The detection completeness for the UDS was computed by Almaini et al. (2014) by first creating a fake UDS background image by removing all the real sources and replacing them with background pixels. Next, for each magnitude of interest, real galaxies which are slightly brighter than the magnitude being tested were rescaled to the magnitude of interest. These artificially dimmed galaxies were then inserted into the fake background and \textsc{sextractor} is run. The completeness is the fraction of correctly extracted sources.\\
\noindent {\bf C-UDS and GOODS-S:} The C-UDS and GOODS-S completeness comes from catalogues of simulated galaxies. These artificial sources had been inserted into the C-UDS or GOODS-S background using the \textsc{iraf} routine \textsc{mkobjects}. In the C-UDS the fake galaxies had either exponential or de Vaucouleurs profiles, whereas in the GOODS-S simulated catalogue the sources had varying ellipticity. In both fields sources had a randomly assigned $H_{160}-$band magnitude, and a distribution of sizes. To compute the completeness fraction, simulated galaxies were randomly selected, in the $H_{160}-$band magnitude bin of interest, such that the distribution of sizes of the simulated galaxies matches that of real galaxies. The distribution of sizes of real galaxies for the C-UDS and GOODS-S was taken from \citet{Vand12}. The fraction of these artificial sources recovered by \textsc{sextractor} was then calculated in each of our magnitude bins. 

To correct our number counts each galaxy was weighted by 1/completeness according to the $H_{160}-$band magnitude of the galaxy. Given that the fake sources created for the UDS completeness were inserted into a sourceless background, these simulation do not take into account confusion. However, the CANDELS completeness simulations do take confusion into account and there is no difference is the galaxy SMFs produced from these two methods of performing completeness corrections. In particular, a difference would be expected at the low mass end and, as none is seen, we conclude that this difference in completeness calculation does not affect the results in this work.

The GOODS-S observations are covered by three different regimes of varying depth. To ensure that each galaxy makes use of the deepest possible near-IR photometry, the GOODS-S data products were taken from a 'max depth' image which varies in depth. The three regions are wide, deep (which includes the ERS) and the HUDF. To account for this, we computed three different completeness curves and weighted each galaxy depending on where it lies within GOODS-S.

Part of the C-UDS is shallower than the deepest region that the simulations used to compute the corrections are based on. \citet{Gala13} showed that the eastern third has a limiting magnitude of $H_{160} <27.90$, whereas the remaining two thirds has a limiting magnitude of $27.90<H_{160}<28.26$. Therefore, some proportion of the galaxies will not be fully corrected and, if this was a large effect it would result in a faint end slope which is shallower than the true value. However, the shallow region of the field is less than one third of the total area and hence the number densities will be dominated by galaxies which are in the deepest part of the field, and therefore the variations in depth will not affect our results drastically. Furthermore, our results match well with results from the literature (see Section \ref{sec:totgsmf}) which utilise samples which are complete to low stellar masses. This suggests our conclusions are not being affected by the C-UDS depth variations. 

The completeness correction allows us to better constrain the low-mass end of the galaxy SMF. However, this correction needs to be applied carefully as overcorrection would lead to an artificially steepening of the low mass slope. In this work the corrections applied to the lowest mass number densities are typically of the order of 10\% in all fields.

For the UDS we use the stellar mass completeness limits computed in \citet{Hart13} and for the C-UDS and GOODS-S we compute the stellar mass limit using the same method. In brief, we take a small redshift slice around the redshift of interested ($\Delta z=0.1$) and take the faintest 20\% of galaxies in this slice. We then rescale the stellar masses of these galaxies to the stellar mass at the 5$\sigma$ limit of the survey ($K$=24.3 for the UDS and $H_{160}=$26.0 for C-UDS). We take the 95$^{th}$ percentile of the scaled stellar mass distribution as our 95\% completeness limit. For a full discussion of this method see \citet{Pozz10}.

\subsection{Environments}
\label{sec:env}
The environments used in this study are are computed by \citet{Lani13} and a detailed discussion on how they are measured can be found there. In brief, the environments are computed from counts in a physical aperture. In \citet{Lani13} a cylinder, of projected radius $400\, \rm kpc$ and depth $1\, \rm Gyr$, was constructed around the galaxy for which the density is measured. A depth of $1\, \rm Gyr$ is substantially greater than the $1\sigma$ measured uncertainty on the redshifts, and hence was large enough to minimise the exclusion of sources due to large photometric redshift  uncertainties, while also being small enough to avoid diluting the number of galaxies in the cylinder.

The number count of real galaxies ($N^{Aper}_{g}$) was normalised to account for edges and holes within the UDS and the final equation for the density is:
\begin{equation}
\rho_{aperture}= \frac{N^{Aper}_{g}}{N^{Aper}_{Mask}} \times \frac{N^{Tot}_{Mask}}{N_{z}}
\label{eq:enviro}
\end{equation}
\noindent where $N^{Aper}_{Mask}$ is the number of pixels within the aperture which were not masked, $N^{Tot}_{Mask}$ is the total number of non masked pixels in the UDS and $N_{z}$ is the number of galaxies in the whole field which lie in the considered $1\, \rm Gyr$ depth.

To divide the systems used in this study into different environment bins, we adopted the same selection method as \citet{Lani13}. For each redshift bin, the mean and standard deviation of galaxies of galaxy environments was calculated for systems above the stellar mass limit. The values of these can be found in Table \ref{tab:denscuts}. Galaxies which live in high density environments were defined as those which had a density greater than one standard deviation from the mean density. The galaxies considered to be in low densities were those which have densities of less than one standard deviation from the mean.

As photometric redshifts are used to measure environments in this study, which can have large errors associated with them, galaxies can be scattered in and out of their true density bin. This problem is particularly worrying in the highest density bin where there are fewer objects and any objects will be preferentially scattered to lower densities. Therefore, we caution against comparing redshift bins to investigate evolution due to redshift. However, in \citet{Lani13} the environments used in this work were found to recover known trends such as the colour-density relation. Furthermore, the location of objects in the highest densities correlates to the positions of known clusters even for faint systems. What this means is that for a given high or low density bin the overall real density distribution will be skewed to the high and low end respectively. Thanks to the excellent number statistics within the UDS this means that we can compare galaxies within a single redshift bin and infer in a statistical manner whether any differences in galaxy properties are correlated with our measure of environment.

\begin{table*}
\begin{tabular}{ | c | c | c | }
\hline
  Redshift Range & $\langle \rho \rangle$ & $\sigma_{\rho}$ \\
\hline
  $0.3<z<0.5$ & 1.20 & 0.57 \\
  $0.5<z<1.0$ & 1.20 & 0.61 \\
  $1.0<z<1.5$ & 1.13 & 0.48 \\
  $1.5<z<2.0$ & 1.11 & 0.46 \\
  $2.5<z<3.0$ & 1.12 & 0.49 \\
  $2.5<z<3.0$ & 1.12 & 0.45 \\
\hline
\end{tabular}
\centering
\caption{The mean environment density ($\langle \rho \rangle$) and standard deviation of the environment density distribution ($\sigma_{\rho}$) in each redshift bin for systems above the stellar mass limit.}
\label{tab:denscuts}
\end{table*}

\subsection{Structural Parameters}
\label{sec:structparams}
\subsubsection{S\'ersic Index}
\label{sec:sersic}
We employ structural parameters from the C-UDS and GOODS-S to construct the galaxy SMF of different galaxy populations. We utilise the S\'ersic indices measured by \citet{Vand12} from the $H_{160}$ C-UDS image using \textsc{GALFIT} (\citealt{Peng10gal}). \textsc{GALFIT} computes the best-fitting S\'ersic model, the parameters of which are, total magnitude, half light radius, S\'ersic index, axis ration and position angle. \citet{Vand12} estimated random uncertainties by comparing independent measures of the same parameters for the same object, in deeper and shallower imaging and reported that the S\'ersic index can be measured to an accuracy of 10\% or better for galaxies brighter than $H_{160} \sim 24.5$.

Our number density estimates extend down to faint magnitudes, where the structural parameter measurements are the least robust. However at a redshift of $z=3$ ($z=1$) the average stellar mass of galaxies with a magnitude of $H_{160} \sim 24.5$ is $M_{*}\sim 10^{9.5}M_{\odot}$ ($M_{*}\sim 10^{8.8}M_{\odot}$). We can therefore still explore the structural parameters of galaxies at these stellar masses to an accuracy of 10\%, despite the difficulties involved with these galaxies being faint.

\subsubsection{CAS Parameters}
\label{sec:cas}
The CAS parameters provide information on the structure of a galaxy. Asymmetry is a measure of the presence of disturbed features in a galaxy. A high asymmetry value indicates a peculiar or merging system. An object with a high concentration parameter has a centrally concentrated light profile, such as a spheroidal system. The clumpiness parameter indicates how smooth the light profile of a galaxy is. For this work, we ran the CAS (concentration, asymmetry and clumpiness) code on the $H_{160}-$band C-UDS and GOODS-S images. In this work the clumsiness parameter was not used as it is found to be the least robust at high redshift due to issues resolving small clumps in these systems with WFC3 (\citealt{Cons03}). Furthermore, the concentration parameter was not used as part of this analysis there for only a description of the asymmetry parameter is included below. For a full description of the CAS code see \citet{Cons03}.

The asymmetry parameter was found by first placing a 3x3 grid over a galaxy. Then a 180$^{\circ}$ rotated image of a galaxy was subtracted from the original image. A background subtraction was included. The equation for this is as follows
\begin{equation}
A = {\rm min} \left(\frac{\sum |I_{0}-I_{180}|}{\sum I_{0}}\right) - {\rm min} \left(\frac{\sum |B_{0}-B_{180}|}{\sum I_{0}}\right)
\label{eq:asym}
\end{equation} 
where $I_{0}$ is the original image pixels and $I_{180}$ is the image after the 180$^{\circ}$ rotation. $B_{0}$ and $B_{180}$ are the values used for the background subtraction. In this equation, the minimum refers to the minimum of the different regions of the grid. The minimum is found by changing the centre around in a 3x3 grid, so that the asymmetry is calculated 9 times at each position. If the minimum of the 9 is the central one then nothing further is done and that value is taken as the asymmetry. If this is not the case it moves the centre to the minimum position and recomputes the asymmetry in the 3x3 grid and continues to do so until the minimum is at the centre.

Uncertainties on the asymmetry are the standard deviation of the background subtraction used in CAS along with counting  uncertainties from the galaxy itself. For a detailed discussion of how the CAS parameters are computed, including centering, measurement of radii and background subtraction see \citet{Cons00} and \citet{Cons03}.

One difference between how CAS parameters have been calculated previously and how we calculated them in this work is the use of median asymmetries. For each galaxy, the asymmetry value is calculated four times, each time using a different area of sky for the background subtraction. We then took a median value as the final parameter. By calculating these structural parameters this way, we were less affected by areas of sky which did not well represent the background value of the image, for example, an area of bad pixels.

Despite the high resolution data used to obtain the CAS parameters, there are still redshift-dependent resolution issues which need to be considered when measuring structural parameters. For example, if asymmetric structures such as clumps or spiral arms are being washed out this will cause the asymmetry values to be too low.  The asymmetry values were tested by artificially redshifting galaxies in the local Universe to redshifts of $z=1.25$, $z=1.75$, $z=2.25$, and $z=2.75$, and then remeasuring their CAS parameters as a function of redshift. As we moved to higher redshifts the asymmetry value systematically decreased. A correction was therefore applied based to these simulations of $\delta A=0.02, 0.04, 0.06,$ and $0.08$ at redshifts $z=1.25$, $z=1.75$, $z=2.25$ and $z=2.75$ respectively. The method to artificially redshift these galaxies is described in full in \citet{Cons03} and \citet{Mort13}.

\section{Results}
\subsection{The Total Galaxy Stellar Mass Function}
\label{sec:totgsmf}
\begin{figure*}
\centering
\includegraphics[width=\textwidth]{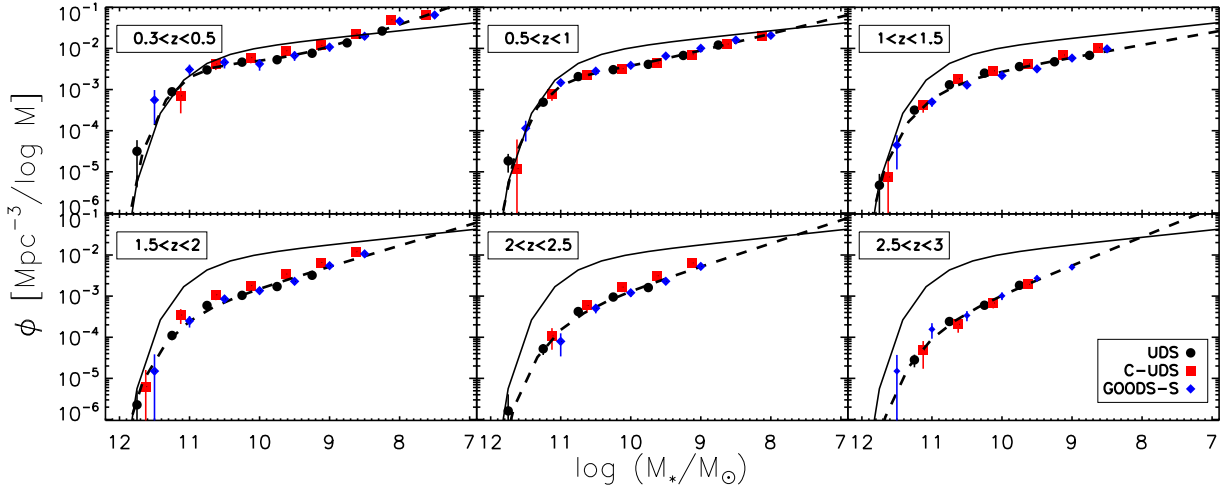}
\caption[The total galaxy SMF for both the UDS and the C-UDS.]{The total galaxy SMF for the combined fields. The black circles are the number densities for the UDS field, the red squares are the number densities for the C-UDS, and the blue diamonds are the number densities for the GOODS-S. The black dashed curves are the Schechter function fits to the combined data. In the redshift range $0.3<z<1$ the fits are double Schechter fits, whereas at redshifts of $z>1$ the fits are single Schechter fits. The solid black curve is the is the local galaxy SMF from \citet{Cole01} converted to a Chabrier IMF.}
\label{fullcanmf}
\end{figure*}

The galaxy SMF is a useful tool for investigating the build-up of stellar mass, and understanding the shape of the galaxy SMF is an important question in astronomy. Many studies have emphasised the importance of understanding the low mass slope which gives us information on the faintest galaxies which are also the most abundant. This slope, and how it evolves with redshift, is difficult to determine robustly due to problems associated with observing faint galaxies. In this paper we use the deep data from C-UDS and GOODS-S to fit the stellar mass function to stellar masses as low as $M_{*}=10^{9}M_{\odot}$ at redshifts of $z=3$.

The UDS benefits from having deep near-infrared data over a very large spatial area, and hence we can construct the galaxy SMF with excellent number statistics. This data set is ideal for constraining the number densities of rarer, massive galaxies, which would be difficult with only the small area of the C-UDS and GOODS-S. The combination of the three data sets used in this work allows us to robustly compute the number densities of galaxies over $\sim$5 dex in stellar mass at redshifts of $z\sim1.25$, and $\sim$2 dex in stellar mass at redshifts of $\sim2.75$.

Figure \ref{fullcanmf} shows the total galaxy SMFs for the combined UDS, C-UDS and GOODS-S datasets. The galaxy number counts were corrected for incompleteness as described in Section \ref{sec:compcalc}. To compute the uncertainties on the number densities we performed a Monte Carlo analysis. Each galaxy's stellar mass and redshift were varied along a Gaussian distribution with standard deviation equal to the 1$\sigma$ measured uncertainty of the stellar masses and redshifts (see Section \ref{sec:masses}). The analysis was repeated on 100 simulated redshifts and stellar masses to obtain simulated number counts. The standard deviation of these values represents the uncertainty due to measurement error. The Poisson uncertainty for each of our number counts was computed and added in quadrature to the uncertainty from our simulations. Finally, uncertainties from cosmic variance were calculated from the \textsc{getcv} code (\citealt{Most11}). This code computes a fractional uncertainty due to cosmic variance as a function of both redshift and stellar mass. These were also added in quadrature to our Monte Carlo uncertainties and Poisson uncertainty and this value was taken as the final uncertainty on the number densities.

To give a feel for the contribution to the uncertainty from each source, we take the $M_{*}\sim10^{10.5}M_{\odot}$ stellar mass bin at redshifts of $z\sim0.4$ and $z\sim1.25$. In the UDS, the Poisson error always contributes the least at both redshifts ($\sim$10\%). There is a small increase in the contribution of uncertainties from SED fitting from the redshift $z\sim0.4$ to the $z\sim1.25$ test bins (from 40\% to 50\%). In the two CANDELS fields the lowest contribution to the uncertainty is from the SED fitting, and cosmic variance has a much larger impact, particularly in the highest of the two redshift test bins ($\sim$55\%).

In this study the number densities were computed as the number of galaxies in a given stellar mass bin normalised by the volume in a given redshift bin. For each of our surveys, separate number densities were computed using the volume of each individual survey. The three surveys were combined in such a way that our number densities are a smooth function of stellar mass (i.e the mass bins used in the fitting are offset by $0.25\,$dex between each survey). The combination of the three fields was then fit by creating an array of all the computed number densities in order of the centre of the stellar mass bin. The combined data sets, in the redshift range $1<z<3$ were fit, using a least-squares fit, with a Schechter function (\citealt{Sche76}) of the form:
\begin{equation}
\phi(M) = \phi^{*}\cdot \rm \ln(10)\cdot[10^{(M-M^{*})}]^{(1+\alpha)}\cdot \rm \exp[-10^{(M-M^{*})}]
\label{eq:schc2}
\end{equation}
\noindent where $\phi^{*}$ is the normalisation of the Schechter function, $M^{*}$ is the turn-over mass in units of dex, $\alpha$ is the slope of the low-mass end of the Schechter function and M is the stellar mass in units of dex. However, in the redshift range $0.3<z<1$ it is clear by inspection that the data is better described by a double Schechter function. The form of the double Schechter function is:
\begin{multline}
\phi(M) = \ln(10)\cdot \exp[-10^{(M-M^{*})}]\cdot 10^{(M-M^{*})}\\
\cdot [\phi_{1}^{*}\cdot 10^{(M-M^{*})\alpha_{1}}+\phi_{2}^{*}\cdot 10^{(M-M^{*})\alpha_{2}}]   
\label{eq:doubsch}
\end{multline}
\noindent where M is stellar mass in units of dex, $M^{*}$ is the turn-over mass in dex, $\phi_{1}$ and $\phi_{2}$ are the normalisations of the two Schechter components and $\alpha_{1}$ and $\alpha_{2}$ are the slopes of the two Schechter components. The reduced $\chi^{2}$ of a single Schechter fit compared to a double is 2.95 and 1.91 in the redshift range $0.3<z<0.5$, and 1.94 and 1.59 in the $0.5<z<1.0$ redshift bin.

\begin{figure*}
\centering
\includegraphics[width=\textwidth,scale=1.2]{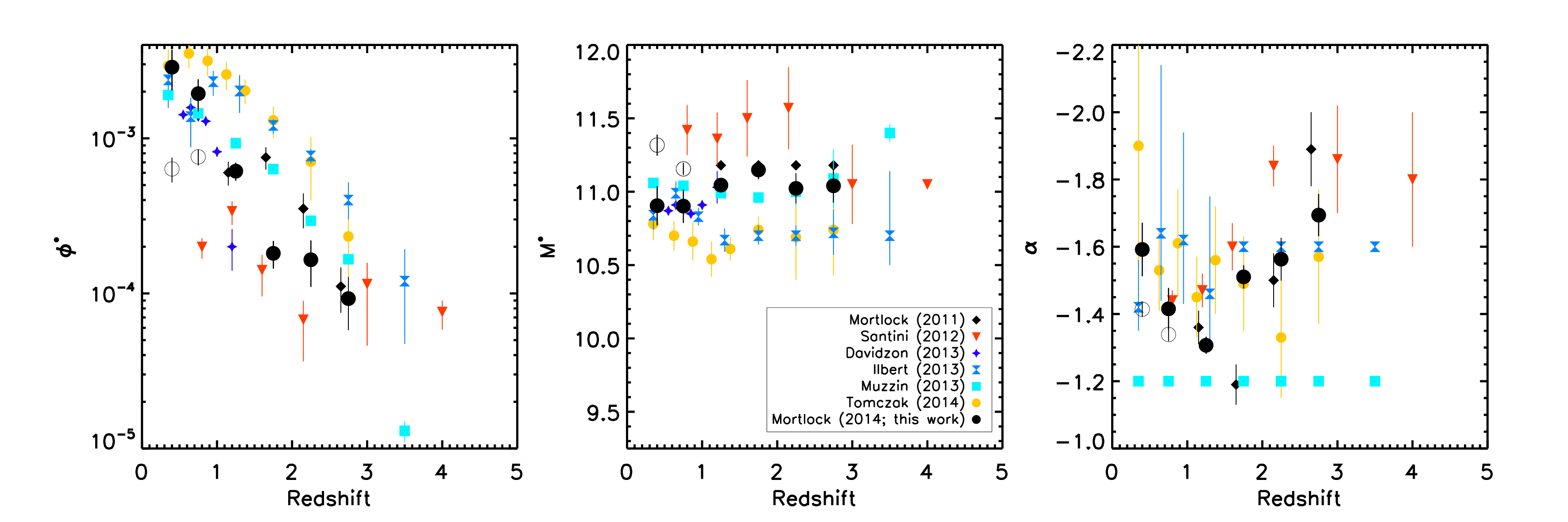}
\caption[The Schechter function parameters.]{The Schechter function parameters of the fits shown in Figure \ref{fullcanmf} (black filled circles). The black empty circle are from the single Schechter fits to the total galaxy SMF (not plotted in Figure \ref{fullcanmf} but shown here for comparison). The black filled diamonds are the results from \citet{Mort11}, the blue filled squares are the results from \citet{Muzz13}, the yellow filled circles are the results from \citet{Tomc14}, the red filled triangles are the results from \citet{Sant13}, the blue filled bow ties are the results from \citet{Ilbe13}, and the blue filled stars are the results from \citet{Davi13}. Where necessary, we converted the literature values of $M^{*}$ to a Chabrier IMF.}
\label{ch2totparams}
\end{figure*}

Figure \ref{ch2totparams} shows the Schechter parameters for the fits to the total SMFs. The Schechter function parameters from the fitting are also in Table \ref{tab:totparamstab} and Table \ref{tab:totparamstabdoub}. An overall increase of $\phi^{*}$ is found from low to high redshift, a steepening of $\alpha$ from low to high redshift, and little evolution of $M^{*}$ across our redshift range.

\citet{Ilbe13} and \citet{Tomc14} fitted their galaxy SMFs with a double Schechter function and hence for $\phi^{*}$ we plot the sum of the normalisations of the two components of their Schechter functions. Furthermore, for $\alpha$, we plot the slope of the low mass component of their double Schechter functions as this is comparable to the low mass slope of a single Schechter function. \citet{Muzz13} fitted with both a double and single Schechter function. Here we plot their results for a single Schechter function for comparison. We discuss the comparisons with other studies in Section \ref{sec:litcomp}.

\subsubsection{Eddington Bias}
\label{sec:eddybias}
When computing the galaxy SMF it is important to consider the effect that the redshift and stellar mass uncertainties have on the shape of the galaxy stellar mass function. The uncertainties in redshift and stellar mass will scatter objects between stellar mass bins. Due to the exponential decline at the high mass end of the galaxy SMF there are fewer objects there. Therefore if an object scatters in or out of a high stellar mass bin this will have a larger effect on the number density. This is known as Eddington bias (\citealt{Eddi13}) and will have the greatest impact on the parameter $M^{*}$.

There have been several previous studies which have explored how Eddington bias changes the shape of the galaxy stellar mass function. \citet{Capu11} found that in the redshift range $3<z<4.25$ the $M^{*}$ parameter is large and that Eddington bias does not have a significant effect. However, at higher redshift ($4.25<z<5$) the high mass end is flattened compared to lower redshifts and this can change the number density of massive galaxies by as much as 0$.$13 dex. \citet{Ilbe13} explore Eddington bias in a similar redshift range to what is considered in this work. By de-convolving their computed number densities by the stellar mass errors they show that across the redshift range $0.8<z<4$ only the high mass turn over is affected and that this change is always less than 0.1 dex.

To understand the effects of Eddington bias on the galaxy SMFs in this study we artificially inflated our stellar masses by adding a number taken from a Gaussian distribution with a FWHM equal to the 1$\sigma$ uncertainty of a given object. The number densities are then recalculated based on these masses and then refit the resulting galaxy SMFs. This was repeated 500 times. By forcing the stellar masses to be larger based on their measured uncertainties a worst case scenario is being imposed in which galaxies are forced into higher stellar mass bins. The high mass turnover $M^{*}$ can increase by as much as to 0.6 dex, although on average is on the order of 0.1 to 0.2 dex.

The uncertainties on the stellar masses used in this study take into account both the uncertainties on the photometry and on the photometric redshifts. Therefore the stellar mass uncertainties incorporate the differences between the individual data sets used and hence our test of Eddington bias takes into account the varying quality of photometry, redshifts and stellar masses. By also testing variations in our fits when inflating the stellar masses we are taking a further step and exploring the highest impact on our results. However, to study the true impact of Eddington bias it is necessary to know the intrinsic galaxy SMF which results in the observed galaxy SMF after all stages of data reduction are taken into account. This requires detailed simulations which we defer to a future paper.
 
\begin{table*}
\begin{tabular}{ | c | c | c | c  | }
\hline
Redshift Range &  $M^{*}$ & $\log \phi_{*}$ & \(\alpha\) \\ 
\hline
$0.3<z<0.5$ & 10.90 $\pm$ 0.13 & -2.54 $\pm$ 0.13 & -1.59 $\pm$ 0.08 \\
$0.5<z<1.0$ & 10.90 $\pm$ 0.11 & -2.71 $\pm$ 0.10 & -1.42 $\pm$ 0.06 \\
$1.0<z<1.5$ & 11.04 $\pm$ 0.04 & -3.21 $\pm$ 0.06 & -1.31 $\pm$ 0.03 \\
$2.0<z<2.5$ & 11.15 $\pm$ 0.06 & -3.74 $\pm$ 0.09 & -1.51 $\pm$ 0.03 \\
$2.0<z<2.5$ & 11.02 $\pm$ 0.10 & -3.78 $\pm$ 0.14 & -1.56 $\pm$ 0.06 \\
$2.5<z<3.0$ & 11.04 $\pm$ 0.11 & -4.03 $\pm$ 0.16 & -1.69 $\pm$ 0.06 \\
\hline
\end{tabular}
\centering
\caption{The single Schechter parameters for the total galaxy stellar mass function. These parameters are plotted in Figure \ref{ch2totparams}.}
\label{tab:totparamstab}
\end{table*}

\begin{table*}
\begin{tabular}{ | c | c | c | c | c | c |}
\hline
Redshift Range &  $M^{*}$ & $\log \phi_{*}^{1}$ & \(\alpha_{1}\) & $\log \phi_{*}^{2}$ & \(\alpha_{2}\)\\ 
\hline
$0.3<z<0.5$ & 10.90 $\pm$ 0.13 & -3.51 $\pm$ 0.30 & -1.59 $\pm$ 0.08 & -2.59 $\pm$ 0.14 & -0.71 $\pm$ 0.31 \\
$0.5<z<1.0$ & 10.90 $\pm$ 0.11 & -3.21 $\pm$ 0.19 & -1.42 $\pm$ 0.06 & -2.93 $\pm$ 0.13 & -0.49 $\pm$ 0.48 \\
\hline
\end{tabular}
\centering
\caption{The double Schechter parameters for the total galaxy stellar mass function. These parameters are plotted in Figure \ref{ch2totparams}.}
\label{tab:totparamstabdoub}
\end{table*}

\subsection{The Galaxy Stellar Mass Function of Blue and Red Systems.}
\label{sec:justcol}
With the combined data set used in this study, we explore the galaxy SMF with respect to colour down to low stellar masses, and with excellent number statistics. The galaxy sample was divided into red and blue galaxies using the $UVJ$ selection described in Section \ref{sec:uvjselec} and number densities were computed in redshift bins between $0.3<z<3$, for the UDS, the C-UDS and the GOODS-S. The colour divided galaxy SMFs were completeness corrected as discussed in Section \ref{sec:compcalc}.

\begin{table*}
\begin{tabular}{ | c | c | c | c | c | }
\hline
  Colour & Redshift Range & $M^{*}$ & $\log \phi_{*}$ & \(\alpha\) \\ 
\hline  $1.5<z<2.0$
Blue & $0.3<z<0.5$ & 10.83 $\pm$ 0.06 & -3.31 $\pm$ 0.05 & -1.41 $\pm$ 0.02 \\
Blue & $0.5<z<1.0$ & 10.77 $\pm$ 0.03 & -3.28 $\pm$ 0.03 & -1.45 $\pm$ 0.01 \\
Blue & $1.0<z<1.5$ & 10.64 $\pm$ 0.02 & -3.14 $\pm$ 0.02 & -1.37 $\pm$ 0.01 \\
Blue & $1.5<z<2.0$ &11.01 $\pm$ 0.06 & -4.05 $\pm$ 0.07 & -1.74 $\pm$ 0.02 \\
Blue & $2.0<z<2.5$ &10.93 $\pm$ 0.07 & -3.93 $\pm$ 0.10 & -1.77 $\pm$ 0.04 \\
Blue & $2.5<z<3.0$ &11.08 $\pm$ 0.11 & -4.41 $\pm$ 0.17 & -1.92 $\pm$ 0.05 \\
Red & $0.3<z<0.5$ & 10.90 $\pm$ 0.05 & -4.87 $\pm$ 0.29 & -1.74 $\pm$ 0.10 \\
Red & $0.5<z<1.0$ & 10.77 $\pm$ 0.03 & -4.11 $\pm$ 0.21 & -1.37 $\pm$ 0.09 \\
Red & $1.0<z<1.5$ & 10.78 $\pm$ 0.02 & -2.96 $\pm$ 0.01 & -0.35 $\pm$ 0.03 \\
Red & $1.5<z<2.0$ & 10.71 $\pm$ 0.03 & -3.31 $\pm$ 0.02 & -0.24 $\pm$ 0.06 \\
Red & $2.0<z<2.5$ & 10.64 $\pm$ 0.04 & -3.55 $\pm$ 0.03 & -0.29 $\pm$ 0.11 \\
Red & $2.5<z<3.0$ & 10.59 $\pm$ 0.06 & -3.78 $\pm$ 0.04 & -0.27 $\pm$ 0.15 \\
\hline
\end{tabular}
\centering
\caption{The single Schechter parameters for the blue and red galaxy SMFs.}
\label{tab:blueredparams}
\end{table*}

\begin{table*}
\begin{tabular}{ | c | c | c | c | c | c | c |}
\hline
Colour & Redshift Range &  $M^{*}$ & $\log \phi_{*}^{1}$ & \(\alpha_{1}\) & $\log \phi_{*}^{2}$ & \(\alpha_{2}\)\\ 
\hline
Red & $0.3<z<0.5$ & 10.90 $\pm$ 0.05 & -4.87 $\pm$ 0.28 & -1.74 $\pm$ 0.10 & -2.80 $\pm$ 0.04 & -0.42 $\pm$ 0.10 \\
Red & $0.5<z<1.0$ & 10.77 $\pm$ 0.03 & -4.11 $\pm$ 0.21 & -1.37 $\pm$ 0.09 & -2.75 $\pm$ 0.02 & -0.27 $\pm$ 0.09 \\
\hline
\end{tabular}
\centering
\caption{The double Schechter parameters for the red galaxy SMFs.}
\label{tab:redparamsdoub}
\end{table*}

\begin{figure*}
\centering
\includegraphics[trim = 4mm 0mm 0mm 6mm, clip,scale=0.55]{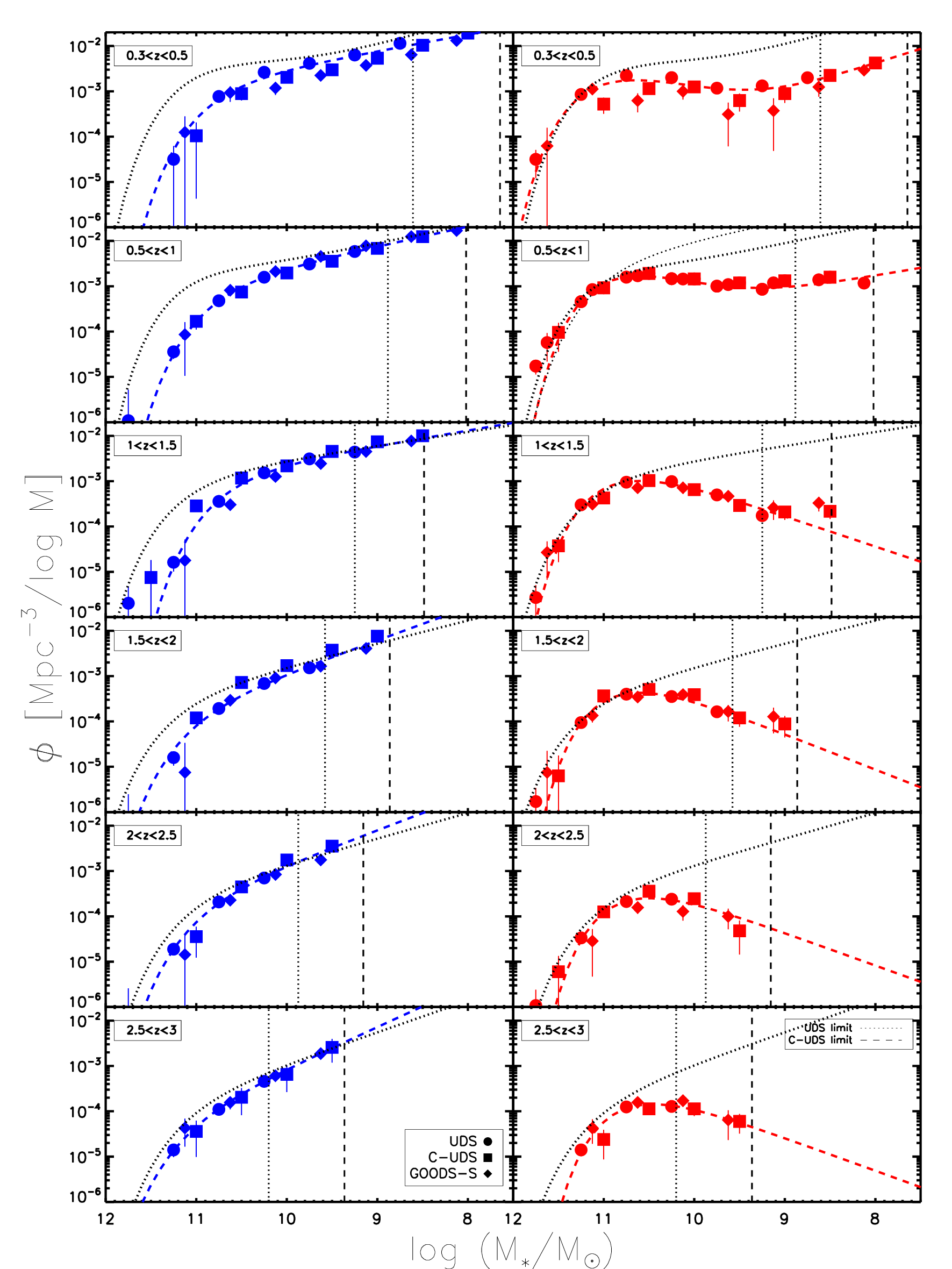}
\caption{The galaxy SMF divided by $UVJ$ selection for the UDS, the C-UDS and the GOODS-S. The left and right panels are the galaxy SMFs of blue and red galaxies respectively. Circles are from the UDS, squares are from the C-UDS, and diamonds are from the GOODS-S. The dashed blue/red lines are the Schechter fits to combined data points. All fits are single Schechter fits with the exception of the red SMFs in the redshift range $0.3<z<1$. The dot/dash vertical lines are the stellar mass limits for the UDS and C-UDS datasets. The dotted black curves are the Schechter fits to the total galaxy SMF, from the relevant redshift bin, in Figure \ref{fullcanmf}.}
\label{coludscanmf}
\end{figure*}

The resulting colour divided galaxy SMFs can be found in Figure \ref{coludscanmf}. All blue SMFs were fitted with a single Schechter functions and the red SMFs in the redshift range $1<z<3$ were also fitted with a single Schechter function. In the redshift range $0.3<z<1$ the red SMFs were fit with double Schechter function. The reduced $\chi^{2}$ of a single Schechter fit compared to a double is 10.22 and 2.40 in the redshift range $0.3<z<0.5$, and 6.85 and 2.04 in the $0.5<z<1.0$ redshift bin. The parameters from the Schechter fits for the blue and red galaxy SMFs are listed in Table \ref{tab:blueredparams} and Table \ref{tab:redparamsdoub}. The uncertainties on the number densities are the result of a Monte Carlo analysis performed in the same way as described in Section \ref{sec:totgsmf}, with an additional Monte Carlo to include the uncertainty in the colours. The uncertainties on the rest-frame magnitudes are a function of redshift, stellar mass and colour. Therefore, by including this additional Monte Carlo simulation we are factoring in changes in the number densities due to possible contamination in the $UVJ$ selection as well as the reliability of the selection as a function of various properties (e.g. the uncertainties will be larger for objects with lower stellar mass). We discuss the blue and red SMFs, and how they compare to previous studies in Section \ref{sec:litcompbr}.

\subsection{The Galaxy Stellar Mass Function In High and Low Densities}
\label{sec:justenv}
The environment in which a galaxy lives plays a key role in its evolution. A galaxy in high density may be subjected to many processes which drive stellar mass build up (e.g. mergers) or affect star formation (e.g. interactions and stripping) more frequently than a galaxy in low density. To investigate how stellar mass builds up in different environments we utilise the environmental information from the UDS (see Section \ref{sec:env} and \citealt{Lani13}).

\begin{figure*}
\centering
\includegraphics[width=\textwidth]{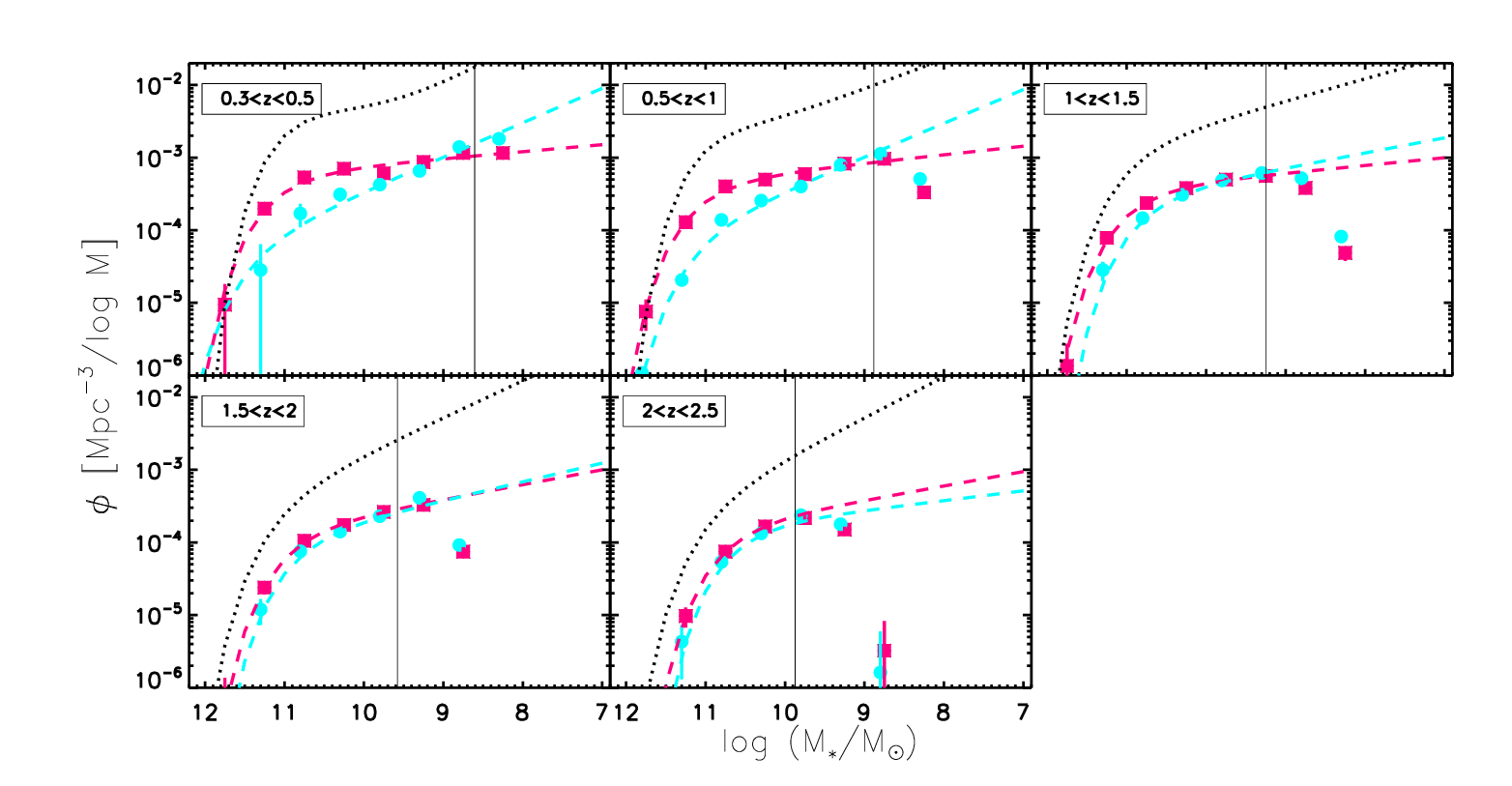}
\caption[The galaxy SMF of galaxies in high and low densities.]{The galaxy SMF of galaxies in high and low densities in the UDS. The blue circles represent galaxies which reside in low densities and the pink squares represent galaxies which reside in high densities (as defined in Section \ref{sec:env}). The dashed lines are the Schechter fits to the data, and the solid black line is the stellar mass limit of the UDS from \citet{Hart13}. The dotted black curves are the Schechter fits to the total galaxy SMF from the relevant redshift bin.}
\label{envudsmf}
\end{figure*}

\begin{table*}
\begin{tabular}{ | c | c | c | c | c | }
\hline
  Redshift Range & Density bin & $M^{*}$ & $\log \phi_{*}$ & \(\alpha\) \\ 
\hline
$0.3<z<0.5$ & High & 11.20 $\pm$ 0.07 & -3.59 $\pm$ 0.07 & -1.10 $\pm$ 0.04 \\
$0.5<z<1.0$ & High & 11.17 $\pm$ 0.05 & -3.70 $\pm$ 0.05 & -1.12 $\pm$ 0.03 \\
$1.0<z<1.5$ & High & 11.06 $\pm$ 0.04 & -3.80 $\pm$ 0.05 & -1.11 $\pm$ 0.03 \\
$1.5<z<2.0$ & High & 11.02 $\pm$ 0.09 & -4.19 $\pm$ 0.11 & -1.21 $\pm$ 0.08 \\
$2.0<z<2.5$ & High & 10.83 $\pm$ 0.14 & -4.14 $\pm$ 0.20 & -1.20 $\pm$ 0.25 \\
$0.3<z<0.5$ & Low & 11.50 $\pm$ 0.65 & -4.55 $\pm$ 0.43 & -1.48 $\pm$ 0.06 \\
$0.5<z<1.0$ & Low & 11.19 $\pm$ 0.18 & -4.38 $\pm$ 0.16 & -1.47 $\pm$ 0.05 \\
$1.0<z<1.5$ & Low & 10.90 $\pm$ 0.08 & -3.91 $\pm$ 0.07 & -1.22 $\pm$ 0.04 \\
$1.5<z<2.0$ & Low & 10.94 $\pm$ 0.13 & -4.29 $\pm$ 0.15 & -1.26 $\pm$ 0.11 \\
$2.0<z<2.5$ & Low & 10.71 $\pm$ 0.20 & -4.15 $\pm$ 0.27 & -1.14 $\pm$ 0.39 \\
\hline
\end{tabular}
\centering
\caption{The single Schechter parameters for the high and low density SMFs.}
\label{tab:hilowdens}
\end{table*}

\begin{figure*}
\centering
\includegraphics[trim = 4mm 0mm 0mm 6mm, clip,scale=0.8]{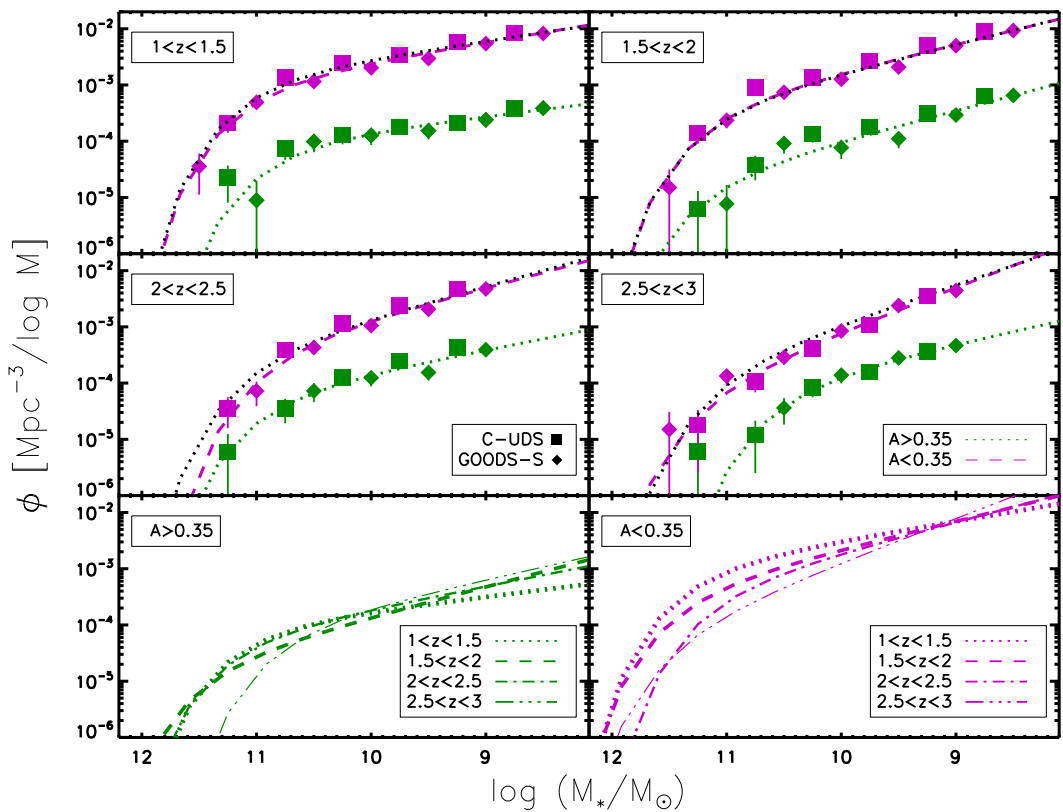}
\caption{The galaxy SMF of the C-UDS and GOODS-S galaxies separated into high asymmetry ($A<0.35$, purple squares and diamonds and purple dashed line) and low asymmetry ($A>0.35$, green squares and diamonds and green dotted line). The dotted black curves are the Schechter fits to the total galaxy SMF, from the relevant redshift bin, in Figure \ref{fullcanmf}. The bottom panels compare the Schechter functions in each redshift bin. All fits are single Schechter fits. The asymmetries are described in Section \ref{sec:cas}.}
\label{asymcanmf}
\end{figure*}

\begin{figure*}
\centering
\includegraphics[trim = 4mm 0mm 0mm 6mm, clip,scale=0.8]{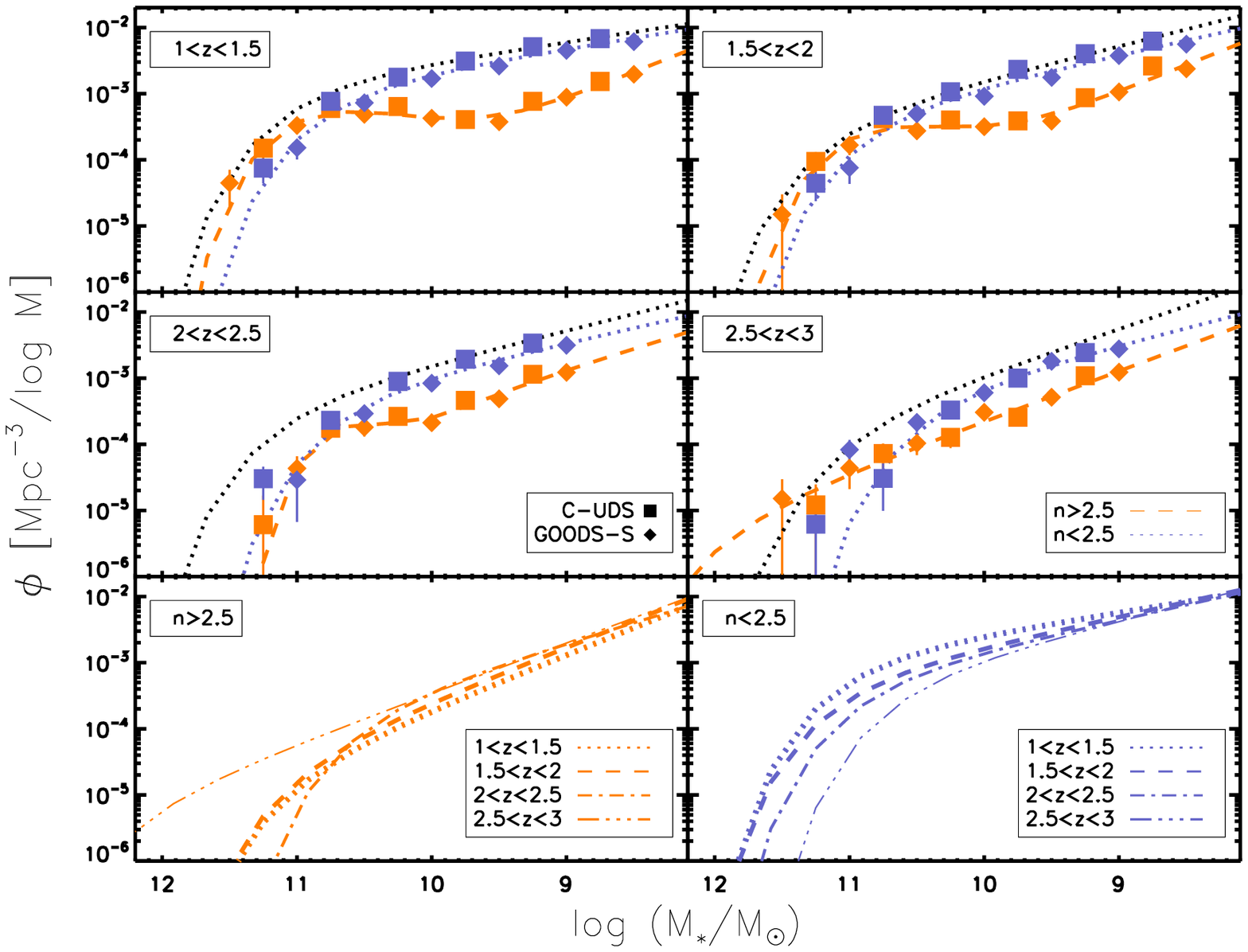}
\caption{The galaxy SMF of the C-UDS and GOODS-S galaxies separated into high S\'ersic index ($n>2.5$, orange squares and diamonds and orange dashed line) and low S\'ersic index ($n<2.5$, blue squared and diamonds and blue dashed line). The dotted black curves are the Schechter fits to the total galaxy SMF, from the relevant redshift bin, in Figure \ref{fullcanmf}. The bottom panels are comparisons of the fits to the high and low S\'ersic indices fits in each redshift bin. All fits to the $n<2.5$ SMFs are single Schechter fits. All fits to the $n>2.5$ are double Schechter fits with the exception of highest redshift bin. The S\'ersic indices are from \citet{Vand12}, see Section \ref{sec:sersic}.}
\label{sersiccanmf}
\end{figure*}

Figure \ref{envudsmf} shows the galaxy SMFs of systems in high and low densities, excluding middle densities (see Section \ref{sec:env} and Table \ref{tab:denscuts}). The SMFs split by environment were completeness corrected as discussed in Section \ref{sec:compcalc}. The uncertainties were computed as in Section \ref{sec:totgsmf}, with an additional bootstrap Monte Carlo analysis to quantify the uncertainties from the environments. This uncertainty was computed by randomly re-sampling the environments and then recalculating the number densities. The result of this was then added in quadrature to the stellar mass and redshift Monte Carlo results to get the final uncertainties.

Our results show little difference in the galaxy SMF of systems in different environments at redshifts of $z>1$. The Schechter fits to the galaxies in dense environments are almost identical to those of galaxies in low densities at these redshifts (see Table \ref{tab:hilowdens} for the Schechter function fit parameters). For redshifts of $z<1$ there is evidence of the high-mass end of the galaxy SMF being more dominated by galaxies in dense environments, although the uncertainties are large. As discussed in Section \ref{sec:env}, the environments in this study can only be used to statistically look at the whole galaxy population. It is possible there is contamination between our high and low density selection and this could be playing a role in our results. One concern is that the contamination will be a function of redshift and stellar mass, hence the shape of the environment SMF could be altered in different ways in different environment bins. In particular at high redshifts, and low stellar masses, the photometric uncertainties will be large and could have more of an impact. To test how this could change the results of this study we take each galaxy and model its redshift uncertainty as Gaussian. Then a new redshift is selected from this distribution 1000 times, and for each new redshift we determine whether or not the object would remain in its original density bin. Then, the average fraction of galaxies which leave their original density bin is calculated as a function of redshift and stellar mass. Using this fraction an upper and lower bound is computed for each number density in Figure \ref{envudsmf}. The lower bound is simply the number density after the removal of the average fraction of galaxies which are removed from their original density bin. The upper bound is taken as the original number density plus the average fraction of galaxies which are removed from the opposite density bin. We find that this upper and lower bound is never larger than the uncertainty on the number densities with the exception of the lowest mass bins in the redshift range $1<z<2$. Here we find that the removal of galaxies based on their photometric uncertainties could produce a slightly lower number density in the lowest density environment. However even taking this into account the high and low density stellar mass functions are still consistent with being the same at low stellar masses. The excess of massive galaxies is thus not affected. The interpretation of these results is discussed further in Section \ref{sec:q2}.

\subsection{The Evolution of the Mass Function for Galaxies with Different Structural Parameters}
\label{sec:structmf}

The SMF of galaxies with different structural parameters gives us insight into the growth of these populations over time, as well as how these objects contribute at different stellar masses. CANDELS provides a deep data set for computing the number densities of galaxies over a large range of stellar masses, as well as high resolution imaging for obtaining accurate structural parameters. Therefore CANDELS is an optimal data set for investigating the SMF of systems with different structural properties. Figures \ref{asymcanmf} and \ref{sersiccanmf} show the galaxy SMF of galaxies from the two CANDELS fields, split by asymmetry and S\'ersic index respectively. The S\'ersic indices are from \citet{Vand12} and the asymmetries were computed using the CAS code as in \citet{Cons03} and \citet{Mort13} (see Section \ref{sec:structparams}). The division in asymmetry is such that galaxies with $A>0.35$ are the most disturbed population (see \citealt{Cons03}). Also, systems with S\'ersic index $n>2.5$ tend to be spheroidal/bulge dominated population and galaxies while $n<2.5$ are the disk/peculiar population (see e.g. \citealt{Lee13}; \citealt{Mort13}; and \citealt{Bruc14}).

The simulations described in Section \ref{sec:totgsmf} were used to correct these SMFs for incompleteness. These simulations contained galaxies with both exponential profiles ($n=1$) and de Vaucouleurs profiles ($n=4$) for the C-UDS galaxies, and with a range of ellipticities for the GOODS-S galaxies. Separate completeness fractions were computed for these two types of galaxies. The completeness fractions for galaxies with exponential profiles were used to correct the $n<2.5$ galaxies, i.e it was assumed the exponential profiles represent low S\'ersic index galaxies. The completeness fractions for galaxies with de Vaucouleurs profiles were used to correct the number densities of $n>2.5$ galaxies, i.e. it was assumed that de Vaucouleurs profiles represent high S\'ersic index galaxies. For GOODS-S it was assumed galaxies with high ellipticity (ellipticity $>$ 0.5), i.e. elongated objects, will have completeness corrections similar to exponential profile objects. For galaxies with low ellipticity (ellipticity $<$ 0.5), i.e. rounded objects, it was assumed galaxies with will have completeness corrections similar to objects with de Vaucouleurs profiles.

Broadly speaking, galaxies with high asymmetries also have low S\'ersic indices, whereas galaxies which are more symmetric have higher S\'ersic indices (e.g. \citealt{Mort13}). Therefore, the same correction was used for the $A>0.35$ ($A<0.35$) number densities as was used for the $n<2.5$ ($n>2.5$) number densities. If we perform the same analysis without the completeness correction (i.e. fitting down to the stellar mass limit), the Schechter function parameters remain virtually unchanged. One particular concern is that the low mass slope will be least robust due difficulties in measuring the structure of low-mass objects. When the Schechter fits were performed without completeness corrections, and only to stellar masses the data is complete, $\alpha$ was changed by typically less than 5\% and always remained within the uncertainty of the value from fitting the completeness corrected SMF. Tables \ref{tab:hilowsersic}, \ref{tab:hisersicdoub} and \ref{tab:hilowasym} list the values of the Schechter parameters for the fits to the structural SMFs. 

Galaxies with low asymmetry dominate at all masses and at all redshifts (over the redshift range $1<z<3$). The evolution of the low asymmetry SMF is very similar to the evolution of the total SMF simply because most galaxies do not have extreme asymmetries by definition. Conversely, the asymmetric population contributes very little to the total mass budget in this redshift range. There is very little evolution in the number densities of these galaxies in our redshift range.

There is a clear differences in the galaxy SMFs of high and low S\'ersic index galaxies. The number densities of low S\'ersic indices galaxies are well fit with a single Schechter function, whereas the number densities of high S\'ersic index galaxies (orange squares and diamonds in Figure \ref{sersiccanmf}) are not.  The reduced $\chi^{2}$ values for the double Schechter fits are 1.53, 1.69, and1.38 compared to 5.45, 5.09, and 3.23 for the single Schechter fits. It has been found extensively in the literature that the total galaxy SMF, and the SMF of red/passive galaxies, show a clear double Schechter function form (\citealt{Dror09}; \citealt{Bald12}; \citealt{Ilbe13}; \citealt{Muzz13}; \citealt{Tomc14}; although to what redshift this persist is still in debate). In this study the same feature is reported in the SMF of galaxies with $n>2.5$. In the highest redshift bin the number densities are closer to the form of a power law, although this could be the result of poor number statistics at the high-mass end. From redshifts $z\sim2.5$ to $z\sim1$ there is rapid evolution at the high-mass end of the SMF of systems with high S\'ersic indices, such that high-mass galaxies with $n>2.5$ appear to be forming quickly.

\begin{table*}
\begin{tabular}{ | c | c | c | c | c | }
\hline
  Redshift Range & S\'ersic Index & $M^{*}$ & $\log \phi_{*}$ & \(\alpha\) \\ 
\hline
$1.0<z<1.5$ &  $n>2.5$ & 11.60 $\pm$ 0.27 & -4.18 $\pm$ 0.20 & -1.30 $\pm$ 0.06 \\
$1.5<z<2.0$ &  $n>2.5$ & 11.74 $\pm$ 0.42 & -4.71 $\pm$ 0.33 & -1.51 $\pm$ 0.06 \\
$2.0<z<2.5$ &  $n>2.5$ & 11.09 $\pm$ 0.24 & -4.45 $\pm$ 0.27 & -1.55 $\pm$ 0.09 \\
$2.5<z<3.0$ &  $n>2.5$ & 11.98 $\pm$ 1.44 & -5.52 $\pm$ 1.26 & -1.76 $\pm$ 0.09 \\
$1.0<z<1.5$ &  $n<2.5$ & 10.78 $\pm$ 0.08 & -3.28 $\pm$ 0.10 & -1.33 $\pm$ 0.05 \\
$1.5<z<2.0$ &  $n<2.5$ & 10.85 $\pm$ 0.10 & -3.61 $\pm$ 0.12 & -1.45 $\pm$ 0.05 \\
$2.0<z<2.5$ &  $n<2.5$ & 10.68 $\pm$ 0.12 & -3.60 $\pm$ 0.17 & -1.46 $\pm$ 0.09 \\
$2.5<z<3.0$ &  $n<2.5$ & 10.42 $\pm$ 0.14 & -3.60 $\pm$ 0.20 & -1.52 $\pm$ 0.11 \\
\hline
\end{tabular}
\centering
\caption{The single Schechter parameters for the high and low S\'ersic indices SMFs.}
\label{tab:hilowsersic}
\end{table*}

\begin{table*}
\begin{tabular}{ | c | c | c | c | c | c | c |}
\hline
Redshift Range & S\'ersic Index &  $M^{*}$ & $\log \phi_{*}^{1}$ & \(\alpha_{1}\) & $\log \phi_{*}^{2}$ & \(\alpha_{2}\)\\ 
\hline  
$1.0<z<1.5$ &  $n>2.5$ & 10.80 $\pm$ 0.12 & -4.93 $\pm$ 0.49 & -1.82 $\pm$ 0.20 & -3.24 $\pm$ 0.09 & -0.32 $\pm$ 0.30 \\
$1.5<z<2.0$ &  $n>2.5$ & 10.78 $\pm$ 0.15 & -4.79 $\pm$ 0.40 & -1.81 $\pm$ 0.15 & -3.49 $\pm$ 0.11 & -0.31 $\pm$ 0.42 \\
$2.0<z<2.5$ &  $n>2.5$ & 10.29 $\pm$ 0.19 & -4.03 $\pm$ 0.33 & -1.63 $\pm$ 0.21 & -3.82 $\pm$ 0.30 & 0.94 $\pm$ 1.13 \\
\hline
\end{tabular}
\centering
\caption{The double Schechter parameters for the high S\'ersic indices SMFs.}
\label{tab:hisersicdoub}
\end{table*}

\begin{table*}
\begin{tabular}{ | c | c | c | c | c | }
\hline
  Redshift Range & Asymmetry & $M^{*}$ & $\log \phi_{*}$ & \(\alpha\) \\ 
\hline
$1.0<z<1.5$ &  $A>0.35$ & 10.86 $\pm$ 0.20 & -4.40 $\pm$ 0.19 & -1.25 $\pm$ 0.09 \\
$1.5<z<2.0$ &  $A>0.35$ & 11.17 $\pm$ 0.35 & -4.99 $\pm$ 0.33 & -1.54 $\pm$ 0.08 \\
$2.0<z<2.5$ &  $A>0.35$ & 10.94 $\pm$ 0.26 & -4.55 $\pm$ 0.28 & -1.40 $\pm$ 0.12 \\
$2.5<z<3.0$ &  $A>0.35$ & 10.51 $\pm$ 0.25 & -4.37 $\pm$ 0.31 & -1.46 $\pm$ 0.16 \\
$1.0<z<1.5$ &  $A<0.35$ & 11.05 $\pm$ 0.08 & -3.31 $\pm$ 0.10 & -1.34 $\pm$ 0.04 \\
$1.5<z<2.0$ &  $A<0.35$ & 11.14 $\pm$ 0.11 & -3.72 $\pm$ 0.13 & -1.50 $\pm$ 0.04 \\
$2.0<z<2.5$ &  $A<0.35$ & 10.86 $\pm$ 0.13 & -3.68 $\pm$ 0.18 & -1.54 $\pm$ 0.08 \\
$2.5<z<3.0$ &  $A<0.35$ & 11.16 $\pm$ 0.24 & -4.36 $\pm$ 0.30 & -1.79 $\pm$ 0.07 \\
\hline
\end{tabular}
\centering
\caption{The single Schechter parameters for the high and low asymmetry SMFs.}
\label{tab:hilowasym}
\end{table*}

\subsection{The Stellar Mass Density}
\label{sec:smd}
\begin{figure}
\centering
\includegraphics[width=0.45\textwidth]{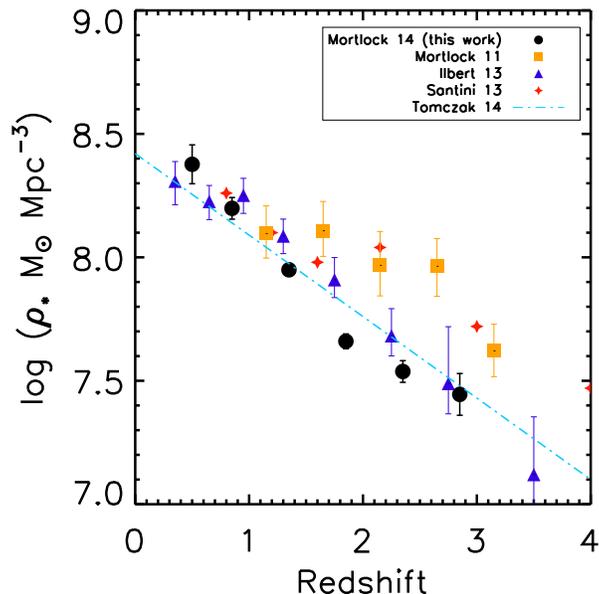}
\caption{The total SMD as a function of redshift. Over-plotted for comparison are various results from the literature (see legend). Where necessary, we converted the literature values of SMD to a Chabrier IMF.}
\label{totsmd}
\end{figure}

\begin{figure*}
\centering
\includegraphics[trim = 15mm 0mm 0mm 6mm,width=\textwidth]{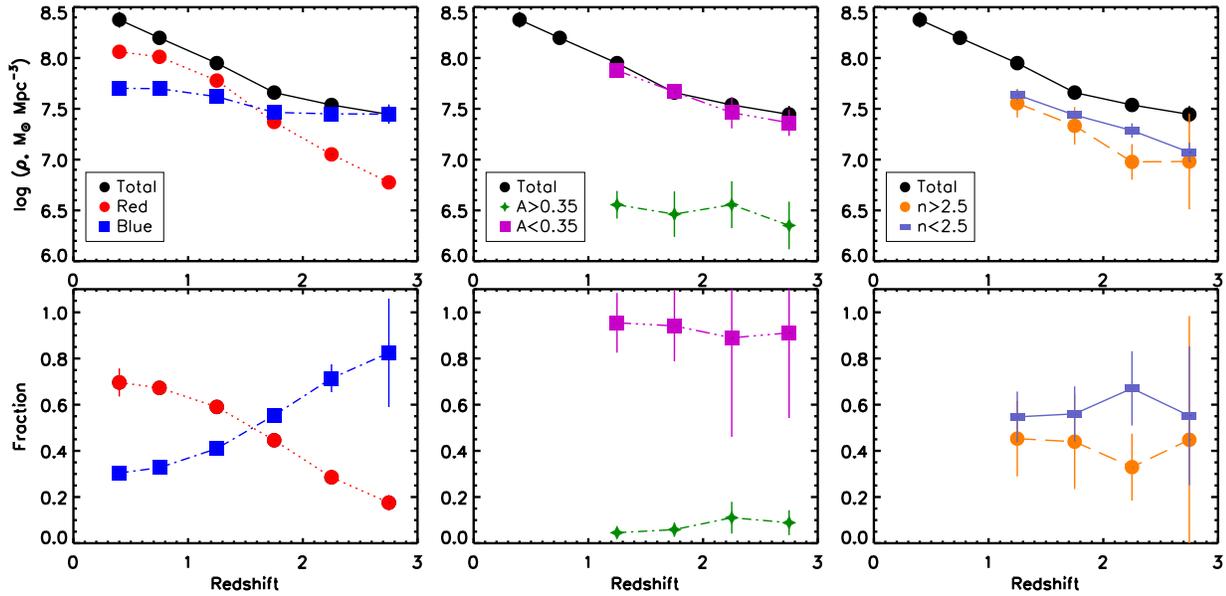}
\caption{Top panels: The stellar mass density as a function of redshift for the total, star forming, quiescent, high and low S\'ersic index, and high and low asymmetry galaxy populations. Bottom panels: The fraction of the total SMD from the total, star forming, quiescent, high and low S\'ersic index and high and low asymmetry galaxy populations. See legend for the meaning of the points and colours.}
\label{allsmd}
\end{figure*}

The stellar mass density (SMD) of the total galaxy population is a useful quantity for investigating the way in which the total stellar mass in the Universe is evolving. Furthermore, by computing the SMD of different galaxy populations we can determine which types of galaxies dominate the stellar mass budget at which epochs, as well as similarities between the growth of different systems. In this study the SMD was computed using the equation
\begin{equation}
\rho_{*} = \int_{M_{*}=7}^{M_{*}=12} M_{*}\times \phi(M_{*}) \mathrm{d}M_{*}
\label{eq:smd}
\end{equation}

\noindent where $\rho_{*}$ is the SMD, $\phi(M_{*}) $ is the Schechter function determined from fitting as discussed in previous sections, and $M_{*}$ is the stellar mass.  Figure \ref{totsmd} shows the total SMD along side several recent literature results. The uncertainties on the SMDs were computed from Monte Carlo shuffling of the Schechter function parameters along a Gaussian with a width equal to the 1$\sigma$ measured uncertainty on that parameter. This was done 100 times and then the SMD for each of these Monte Carlo realisations was computed. The final uncertainty was the standard deviation of these realisations.

A steady increase of the total SMD is shown in Figure \ref{totsmd}. This increase is the build-up of all stellar mass in the Universe. Taking the SMD at redshift $z=0$ from the relation in \citet{Tomc14} we find that at redshift  $z=1.25$, 35\% of the SMD in the Universe has been built up. By redshift $z=0.75$, 58\% of the total SMD had been built up, i.e. $\sim$50\% of the total SMD is formed around redshift of $z=1$. Figure \ref{totsmd} shows that there is a large difference between the SMD from \citet{Mort11} and \citet{Sant13} and the remainder of the SMDs plotted, including the SMD computed in this work. \citet{Mort11} and \citet{Sant13} used data from small area surveys and therefore had difficulty constraining the high-mass end. This could have resulted in a higher value for the high-mass turn over (see Figure \ref{ch2totparams}) and therefore stellar mass would be added to the total SMD at the high-mass end. We therefore argue that the higher values for SMD are a result of poor constraints on the high-mass end of the galaxy SMF.

The SMD for the total, star forming, quiescent, high and low S\'ersic index, high and low asymmetry, and high and low density galaxy populations is also plotted as a function of redshift in Figure \ref{allsmd}. Considering the total galaxy population split into star forming and quiescent galaxies, it is clear the increase in the total population is driven by the rapid growth of the quiescent population. The SMD of the star forming population remains roughly constant over our whole redshift range. Furthermore, there is a clear cross-over around redshift of $z\sim2$ where the quiescent population becomes dominant over the star forming population.

A similar cross-over behaviour is found for the SMD of galaxies with high and low S\'ersic index. At $z>2$ there is not a substantial difference between the SMD of these two types of system. However, there is evidence that the SMD of high and low S\'ersic index galaxies begins to converge at $z<2$.  Several previous studies (e.g., \citealt{Buit08}; \citealt{Bruc12} \citealt{Mort13}) have found a cross-over in the dominance of galaxies with disk and spheroidal morphology (which are traditionally considered to be low and high S\'ersic index galaxies) around this redshift.

For galaxies of low asymmetry, the evolution of the SMD is very similar to that of the total galaxy population. This is not a surprise as high asymmetry is selecting galaxies which are in the most peculiar stages of a merger whereas low asymmetry is selecting the remainder of the galaxy population. Interestingly, the SMD of systems with high asymmetry is roughly constant across our redshift range. This is similar to the behaviour of the star forming SMD (and SMF). This could indicate that, in a similar way to the star forming and quiescent SMF, galaxies move along and/or off the high asymmetry SMF in such a way that the evolution of the high asymmetry SMF is stellar mass independent.

\section{Discussion}
\label{sec:conc}
\subsection{Comparison to the literature}
\subsubsection{The total galaxy SMF}
\label{sec:litcomp}
The Schechter function parameters describe the growth of the total galaxy population ($\phi^{*}$), as well as the evolution of high-mass ($M^{*}$) and low mass galaxies ($\alpha$). From high to low redshift, galaxies are forming and growing in stellar mass hence an increase in $\phi^{*}$ should be seen. The evolution of this parameter has long been established (e.g. \citealt{Font04}; \citealt{Dror05}; \citealt{Borc06}; \citealt{Bund06}; \citealt{Font06}; \citealt{Pere08}; \citealt{Kaji09}; \citealt{Marc09}; \citealt{Ilbe10}; \citealt{Capu11}; \citealt{Mort11}; \citealt{Sant12}; \citealt{Davi13}; \citealt{Ilbe13}; \citealt{Muzz13}; \citealt{Tomc14}; \citealt{Dunc14}), but with field to field variation there have been differences in the absolute value (left hand panel of Figure \ref{ch2totparams}). In this plot the results from \citet{Ilbe13} and \citet{Tomc14} are the sum of two normalisations from a double Schechter fit and so are not directly comparable. It is noted that differences in the form on the Schechter function further highlight the spread in results.

Some of the earlier studies of the galaxy SMF have showed that the value of $\alpha$ is steep and gets steeper as redshift increases (e.g. \citealt{Font04}; \citealt{Font06}; \citealt{Marc09}). However the absolute value of $\alpha$, and at what redshift the evolution continues to can depend greatly on the stellar mass limit of the survey. As technology has advanced and observations of faint galaxies has become more routine it has become clear that $\alpha$ steepens out to redshifts of $z\sim3$ (\citealt{Kaji09}; \citealt{Mort11}; \citealt{Sant12}; \citealt{Tomc14}) and higher (\citealt{Capu11}; \citealt{Sant12}; \citealt{Dunc14}). This is confirmed in this work and indicates that low stellar mass galaxies contribute significantly to the total stellar mass budget in the Universe at these epochs. The flattening of $\alpha$ with time is likely a result of low stellar mass galaxies increasing in stellar mass via merging or some process such as in-situ star formation.

While the parameters $\phi^{*}$ and $\alpha$ show strong evolution in our redshift range, the value of the characteristic turn-over $M^{*}$  remains roughly constant from a redshift of $z=0.3$ to $z=3$ (middle panel of Figure \ref{ch2totparams}). The lack of evolution of $M^{*}$ implies that although the number density of massive galaxies is increasing over time due to the increase in normalisation, the knee of the mass distribution is not evolving. What this means is that the typical stellar masses of systems around the knee of the galaxy SMF are not evolving with redshift. This is also evident when comparing the high-mass end of the galaxy SMF at high redshift to that of the local Universe. The number densities of the most massive galaxies ($M_{*}>10^{11.5}M_{\odot}$) evolve very little from $z\sim3$ to $z=0$. Furthermore, galaxies with stellar masses of  $M_{*}>10^{11}M_{\odot}$ have reached their local number density between redshifts of $z=1-2$. The lack of evolution is consistent with downsizing (e.g. \citealt{Cow96}), where the most massive galaxies form their stellar mass first.

There is a huge amount of work in the literature dedicated to studying the galaxy SMF, however, we choose to compare to the most recent literature results here, which also probed similar redshift range to this study. The main motivations for this is to compare to studies which either are a) complete to similar stellar masses as in this work for a good comparison to $\alpha$ or b) from large area surveys for a good comparison to $M^{*}$. Furthermore, by being selective in our comparison we are able to compare to studies which fit to similar stellar population models (BCO3 models, with exponentially declining star-formation histories). The only differences which could cause problems are the inclusion of emission lines and differences in the IMF. \citet{Mort11}, \citet{Sant12} and \citet{Muzz13} use a different IMF to this study however this is always accounted for when comparing results. Finally, \citet{Ilbe13} and \citet{Tomc14} include nebular emission lines when computing galaxy properties however as noted in Section \ref{sec:masses} the presence of nebular emission has little impact on this work.

Although there is good agreement on the evolution of the Schechter function parameters there are generally differences in the absolute values as shown by the comparison of the literature results in Figure \ref{ch2totparams} (see legend). In terms of $M^{*}$ and $\alpha$ the biggest differences are likely the result of the different depths and areas covered by surveys. Large differences in $M^{*}$ will results from poor constraint of the high-mass end of the galaxy SMF in surveys with small areas. This is supported by the fact that the results from the two smallest areas, \citet{Mort11} and \citet{Sant12}, find the highest values for $M^{*}$. In this work the large area of the UDS is utilised to constrain $M^{*}$. Comparing our results for $M^{*}$ to \citet{Muzz13}, who used a comparable survey area and fit with a single Schechter function (hence the parameters are directly comparable), we find excellent agreement across out whole redshift range. \citet{Davi13} construct the galaxy SMF over the much larger area of VIPERS. This allows for excellent constraint on the high-mass turn-over and where we overlap in redshift our values of $M^{*}$ differ by at most $\sim$0.1 dex.

Often, survey area is compromised for depth, as having a deep survey is key for constraining the low mass slope $\alpha$. In this work, the CANDELS data allowed us to fit to low stellar masses, and hence this work improves upon other studies which have only large area data (e.g. \citealt{Ilbe13} and \citealt{Muzz13}). In comparison to \citet{Mort11}, who have the deepest data set which is fit by a single Schechter function, excellent agreement is found for the low mass slope. \citet{Tomc14} have a similar stellar mass limit to this study and fit with a double Schechter function across their whole redshift range. The double Schechter fits to the redshift range $0.3<z<1.0$ (see Section \ref{sec:structmf}) are in good agreement in $\alpha$ at $z\sim1$.

\citet{Tomc14} not only have a similar stellar mass limit to this study, they also combine deep data with wide area data to constrain the high-mass turnover as we do in this work. However, the wide UDS data used here is $\sim0.5$ deg$^2$ larger in area than the NEWFIRM Medium-Band Survery (NMBS) which is crucial for finding the rarest high-mass galaxies. Therefore, in this work we have a better constraint on both $M^{*}$ and $\alpha$ thanks to the combination of both deeper and larger area surveys than previous studies.

\subsubsection{The blue and red galaxy SMFs}
\label{sec:litcompbr}
In this study the total galaxy SMF was divided into blue and red galaxies using the UVJ selection technique. This division was used as a proxy for the star formation activity within a system. The blue, star forming, SMF shows very mild evolution in shape at $z<2$ and little evolution in normalisation. At redshifts of $z>2$ the slope of the star forming SMF steepens, and the high-mass turnover increases (see Table \ref{tab:blueredparams}), however there is very little further evolution. The lack of evolution of the star forming SMF is reflected in the almost constant SMD of star forming galaxies (Figure \ref{allsmd}). This is in agreement with, e.g. \citet{Borc06}; \citet{Bell07}; \citet{Peng10}; \citealt{Mort11}; \citealt{Mous13}); \citet{Ilbe13}; \citet{Muzz13} and \citet{Tomc14}, who have shown that at redshift of $z>2$ the SMD of star forming systems begins to decrease. Furthermore, some studies have found the blue SMF to have a double Schechter function form (\citealt{Dror09}, \citealt{Ilbe13} and \citealt{Tomc14}), however we see no evidence for either of these trends in our study.

The red, quiescent SMF evolves very little in $M^{*}$ whereas the low mass slope becomes steeper with look back time. In our lowest redshift bin, $0.3<z<0.5$, the number density of galaxies with $M_{*}\sim10^{11}M_{\odot}$ has grown by only a factor of $\sim$1.5 since redshift of $z\sim2$, whereas the number density of galaxies with $M_{*}\sim10^{9.5}M_{\odot}$ has increased by a factor of 5. This suggests that the growth of quiescent galaxies mass dependent, i.e. there are two separate high-mass and low-mass populations growing in stellar mass. This result is in agreement with many literature results. The bimodal form of the red SMF was first noted in \citet{Dror09} and has since been shown in the local Universe (\citealt{Bald12}) and out to redshift of $1<z<1.5$ (\citealt{Ilbe13}; \citealt{Muzz13}; \citealt{Tomc14}). The most massive galaxies build-up their stellar mass early in the life of the Universe, and also quenched earlier. However the number density of low mass galaxies build-up rapidly in our redshift range hence low-mass systems are quenching quickly. Furthermore, the normalisation, $\phi^{*}$, increases by almost an order of magnitude from redshift $z=3$ to $z=0.3$, reflecting the rapid build-up of the population of galaxies whose star formation has ceased.

The mass dependent growth of red systems leads to a double Schechter function form to the quenched galaxy SMF in the lowest two redshift bins. There is hint of the double Schechter feature in the $1<z<1.5$ redshift bin, however it is unclear if this feature is present above a redshift of $z=1$ as it  is close to the completeness limit. The double Schechter function form is likely the result of two distinct galaxy populations dominating at different stellar masses. \citet{Peng10} suggested these two quenched populations are either mass quenched or quenched as a result of their environment (see Section \ref{sec:q2}). In this picture, the low-mass end is dominated by environment quenched galaxies, hence the upturn in the quenched galaxy SMF is directly linked to the importance of the environment quenching process. Although we cannot say anything about environment quenching at $z>1$ here, the upturn seems more prominent at $z\sim0.4$ compared to at $z\sim0.75$. This could be an indication that environment quenching is becoming more important in the lowest redshift bin (\citealt{Muzz13}).

The rapidly changing SMF of red, or quenched galaxies, and the lack of evolution of blue or star forming galaxies, is an indication of a balance between the growth of galaxies on the star formation main sequence and the processes which shut off star formation. When systems are quenched, they move off the star forming SMF and onto the quenched SMF. The cessation of star formation is a function of stellar mass, yet this is not reflected in the shape of the star forming SMF. The build up of stellar mass in star forming galaxies must happen in such a way that as galaxies move along the star forming SMF (i.e. grow in stellar mass), the number density of galaxies of all stellar masses is preserved.

The impact of contamination within the $UVJ$ selection has been discussed in Sections \ref{sec:uvjselec} and \ref{sec:justcol}. However, as a final test of how robust to contamination the results discussed here are, a Monte Carlo simulation is performed. In this simulation a given percentage of random objects are moved from the red to blue bin, or vice versa, 500 times. The approach is complementary to the Monte Carlo which produces the uncertainties on the number densities. In the latter case the uncertainty on the number densities is representative of how the $UVJ$ selection impacts these results as a function of, not only redshift and stellar mass, but also of data set. In our second approach we are testing how robust our results are to a certain level of contamination.

When we shift 5\% of objects from red to blue and vice versa this has no impact on the results of this work. When 10\% of objects are shifted to the opposite colour bin we find this does have some impact on the fitted Schechter function parameters of the red SMFs. The low mass slope becomes less shallow at redshift of $z>1$ and $M^{*}$ is increased by $\sim0.2$ dex. However, the change in the absolute value of the Schechter function parameters does not affect any results discussed in this section e.g. the double Schechter function form of the red SMFs and the steepening of $\alpha$ of the red SMF with redshift. Furthermore, the change is not enough to impact the SMD results discussed in Section \ref{sec:smd}. If the percentage of objects shifted between colour bins is changed to 20\% the impact on the Schechter function parameters is more severe. M$^{*}$ is increased by $\sim0.4$ dex and $\alpha$ becomes steep (-1.3 to -0.9) at redshifts of $z>1$. This in turn alters the SMDs by $\sim0.2 - 0.3$ dex. However, at $z<1$ the red SMF is impacted less and still retains the double Schechter function form.

\subsection{The Link Between Quenching and Structure}
\label{sec:q1}
\begin{figure*}
\centering
\includegraphics[trim = 0mm 0mm 0mm 0mm, clip,scale=0.8]{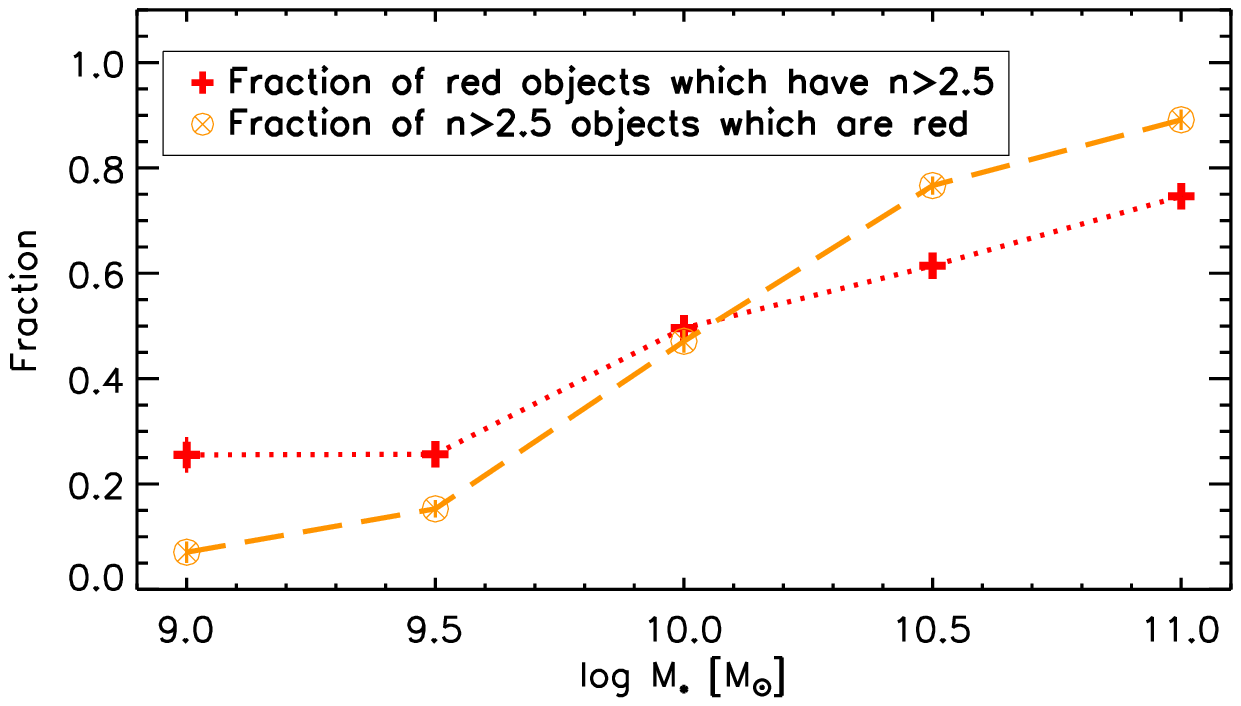}
\includegraphics[trim = 0mm 0mm 0mm 0mm, clip,scale=0.8]{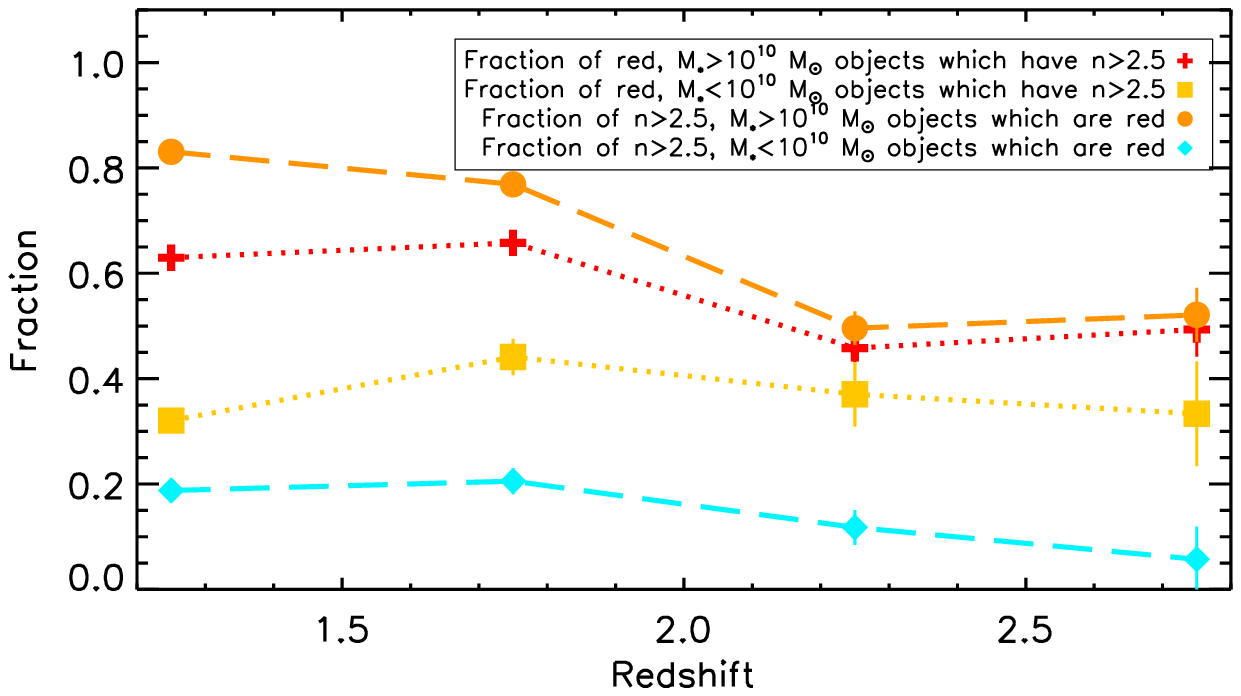}
\caption{The fraction of galaxies which are red and have high S\'ersic index (red crosses joined by dotted lines) and the fraction of galaxies which have high S\'ersic index and are red (orange circles with crosses joined by dashed lines). The top panel shows these fractions as a function of stellar mass. The bottom panel shows these fractions as a function of redshift and split by stellar mass. See legend for symbol descriptions}
\label{UVJsersicfrac}
\end{figure*}

There is debate in the literature regarding the link between passivity and the structure of a galaxy. This is frequently cast in terms of the presence of a bulge correlating with a galaxy being passive (e.g. \citealt{Bell12}; \citealt{Cheu12}; \citealt{Barr13}; \citealt{Fang13}; \citealt{Lee13}; \citealt{Lang14}; and \citealt{Will14}). There are exceptions e.g. the presence of passive disks (e.g. \citealt{Bruc12}; \citealt{Mclur13} and \citealt{Bruc14}) and hence there is still question as to if passivity is caused by the presence of a bulge or vice versa. In this work, similarities are found between the shape of the galaxy SMF of high S\'ersic index galaxies (i.e bulge/spheroidal galaxies at these redshifts, see \citealt{Lee13}; \citealt{Mort13}; and \citealt{Bruc14}) and the red, or quenched, SMF. It is also shown in this study that the SMDs of red systems and $n>2.5$ systems are comparable. This suggests a  similar amount of stellar mass is tied up in red galaxies and $n>2.5$ galaxies across our redshift range.

To discern if these two populations are linked we plot the fraction of galaxies which are red and have high S\'ersic indices, and the fraction of high S\'ersic index galaxies which are red, as a function of stellar mass and redshift (see Figure \ref{UVJsersicfrac}). The uncertainties in this plot were computed from varying the uncertainties on the redshifts, stellar mass, colours and S\'ersic index with standard deviation equal to the 1$\sigma$ measured uncertainty of each property. This Monte Carlo analysis was done 100 times, and the fraction was recalculated for each run. The final uncertainties on the fractions in Figure \ref{UVJsersicfrac} were the standard deviations of these Monte Carlo runs. There is no evidence of a trend with redshift for red galaxies with high S\'ersic index. For high S\'ersic index galaxies which are red, the fraction increases as a function of redshift for both stellar masses. In the top panel of Figure \ref{UVJsersicfrac} the relationship between a galaxy being red and having high S\'ersic index is a strong function of stellar mass. Of the galaxies with $n>2.5$ and stellar mass $M_{*}\sim10^{11}M_{\odot}$, $\sim90\%$ have red colours, and $\sim75\%$ of red galaxies have $n>2.5$. These percentages drop to roughly 25\% for red galaxies with $n>2.5$, and almost nothing for the high S\'ersic index, red galaxies, when considering low-mass systems.

Our results suggest that, if indeed high S\'ersic index links to the cessation of star formation, this is only true in massive galaxies and not in low mass systems. This is supported by the \citet{Deve09} who construct the $K$-band luminosity function and show that bright/faint systems are dominated by elliptical/disk morphologies.These results are indicative of mass dependent processes at play. A possible scenario is that for high-mass galaxies some process (or processes) act on a galaxy and instigate the formation of a bulge and also quenches that system. In low mass galaxies bulge formation does not link to the cessation of star formation. We cannot discern from these results if quenching and bulge formation happens simultaneously or not in massive galaxies. However we suggest a possible link to feed-back from black holes. The mass of a black hole scales with the mass of a galaxy and the bulge (e.g. \citealt{Ferr00} and \citealt{Gebh00}) and hence feed-back from a massive black hole could result in the formation of a bulge whilst also being powerful enough to shut off star formation. However, our results suggest that in a less massive galaxy the process responsible for bulge formation, e.g. the black hole feed-back, would not be strong enough to remove/cut off a galaxies gas supply allowing star formation to continue.

\subsection{The Link Between Quenching and Environment}
\label{sec:q2}
\begin{figure*}
\centering
\includegraphics[trim = 20mm 115mm 0mm 10mm, clip,scale=0.4]{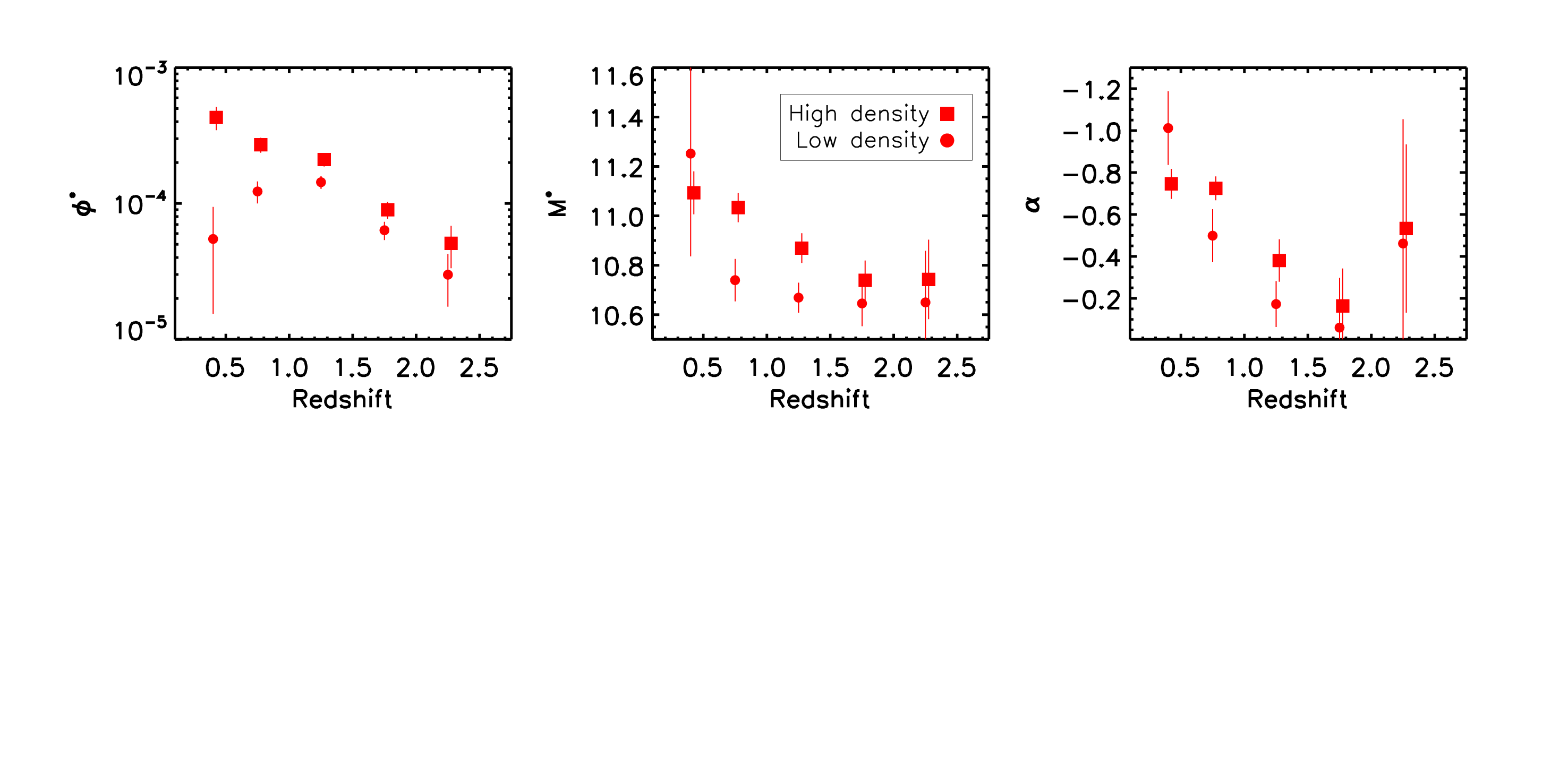}
\caption{The Schechter function parameters from fits to the number densities of red, high density (red squares) galaxies, and red, low density (red circles) galaxies.}
\label{envcolparams}
\end{figure*}

A picture is beginning to emerge regarding the role environment plays in quenching galaxies. In the local Universe the fraction of quenched galaxies is high for massive systems, or if a galaxy lives in high density environments (\citealt{Bald06} and \citealt{Peng10}). This suggests two separate 'mass' and 'environment' quenching processes. These two processes are separable, and that the evolution of the galaxy SMF of blue and red galaxies can be explained with these two processes even out to high redshift (e.g., \citealt{Ilbe13} and \citealt{Scov13}). There is also discussion in the literature regarding a comparable theory of separate quenching processes (e.g. \citealt{Deke14}; and \citet{Woo14}). These studies suggest quenching channels which depend on halo mass and structure. The halo mass channel is supported by observations of differences in passivity in central and satellite galaxies (e.g. \citealt{Woo14}), and also supported by theory (e.g. \citealt{Deke06}). The relationship between quenching and structure is discussed in detail in Section \ref{sec:q1}.

To explore the role that environment plays in quenching and shaping the galaxy SMF, we utilised the environments in the UDS field (see Section \ref{sec:env}) to compute the number densities of red galaxies in the high and low density environments. These number densities were fit with a Schechter function, the parameters for which are shown in Figure \ref{envcolparams}.  At $z>1.5$ the two galaxy SMFs are consistent with one another. The uncertainties are large, but environment will have a weaker impact on galaxies at higher redshifts as the densest structures have yet to form, hence this is expected and is consistent with Figure \ref{envudsmf} where the two galaxy SMFs are identical at $z>1.5$.

Thanks to the excellent number statistics in the UDS we can still constrain the high-mass end of these SMFs even after this division. However, the difficulty here lies in the shallower depth of the UDS (e.g. compared to the CANDELS data used in this work) which creates problems when drawing conclusions from the low mass slope. The uncertainties on $\alpha$ are large, therefore it can only be tentatively suggested that, at $0.5<z<1.5$, Figure \ref{envcolparams} hints at a steeper low mass slope in high density, compared to low density. Again, it is stressed that the uncertainties mean we can say little more about $\alpha$, however what is hinted at by the data does make sense in the context of environment quenching. The total quenched SMF shows a steep faint end slope (right hand panel of Figure \ref{envcolparams}), and in \citealt{Peng10} it is suggested that low mass satellite galaxies are the cause of this. These systems are starved of their star formation fuel by e.g. gas stripping or harassment in high density and hence we would expect a steeper faint end slope in higher density environments.

In the middle panel of Figure \ref{envcolparams} the high-mass turnover is clearly higher for the galaxies which are, on average, in higher densities in the redshift range $0.5<z<1.5$. The larger values of $M^{*}$ are indicative of high density causing the build up of high-mass, red galaxies. Figure \ref{envudsmf} shows that the number densities of the total high-mass population in high densities are larger than in low density environments. In this study it is shown that this is also true for the quenched galaxy population. The fact there is an excess of quenched galaxies in high densities, not just in the total population of galaxies in high densities, links environment to the cessation of star formation in a galaxy. This increased quenching efficiency could be linked to processes which occur more frequently in high density environments, such as mergers or mass quenching due to hotter halos.

One caveat of this work is that photometric redshifts are used for computation of environments, and hence the contamination between low and high density is higher than ideal for this test. The same test as in Section \ref{sec:justenv} is performed on the red, high and low density SMFs. When adjusting the number densities for the possible fraction of galaxies lost/gained in a given density bin due to uncertainties on the photometric redshifts we find no impact on our conclusions regarding $M^{*}$. However, the value of $\alpha$ for the high density, red SMF, in the redshift bin $1<z<1.5$ is reduced and is consistent with the value of $\alpha$ for the low density red SMF. This further stresses the difficulty with drawing any conclusions from the low mass slope of quenched galaxies in different environments. Furthermore, contamination from photometric redshift errors means we cannot say for sure that galaxies in our high density bin are quenched truly as a result of the environment which they live in. We therefore caution that the results found here are tentative but, due to the unique environmental information available in the UDS across such a large area and redshift range, are a first step in understanding the links between environment and quenching. 

\subsection{Low Stellar Mass Galaxies in High Densities}
\label{sec:lowmdens}
\citealt{Cons02}; \citealt{Penn08}; \citealt{Vulc11}; and \citealt{Vulc13} explored the local galaxy SMF and luminosity function with respect to environment . These studies do not agree as to whether low stellar mass galaxies (or faint galaxies) are more abundant in high densities (i.e. do high densities form low stellar mass galaxies faster when considering the total galaxy population). There is no excess of low stellar mass galaxies at high densities in our results, which suggests this environmental effect is not present. This is the case at all redshifts. To explore this further would require a deeper data set to more accurately determine $\alpha$.

To test how different definitions of high and low density may impact our results, the division between high and low density was altered in several ways. Firstly, the same selection method as previously described was used, but with respect to log $\rho$, rather than $\rho$, so that the distribution of densities is closer to being Gaussian. Secondly, galaxies were selected in the upper and lower 16\% of the distribution so that each density bin contained the same number of galaxies, and finally a constant density cut in each redshift range was tested. For the constant density cut we used the mean and standard deviation of the densities of all galaxies at all redshifts, rather than in separate redshift bins as in Figure \ref{envudsmf}. Varying our definition of density in these ways had no impact on our results. We keep in mind a cautionary note regarding possible contamination between the high and low density bins. It is possible similarities between the two SMFs are a result of the difficulties in clearly defining environment and improved photometric redshifts from future surveys will provide a clearer insight into this in the future.

\section{Summary}
\label{sec:summary}
Using a combination of the UDS, CANDELS UDS and CANDELS GOODS-S data sets, the galaxy stellar mass function for the total galaxy population over the redshift range $0.3<z<3.0$ is constructed. These data sets provide a combination of extremely deep near infra-red data and excellent number statistics. Each of the data sets used here give us the power to accurately probe different parts of the galaxy SMF, as well as the galaxy SMF with respect to different galaxy properties. We take advantage of the environment information available in the UDS to investigate how the form of the SMF is affected by environment. Also utilising the high resolution imaging available in the C-UDS and GOODS-S fields, the galaxy SMF with respect to both S\'ersic index and asymmetry is explored.

The major results of this paper are as follows:
\begin{itemize}
\item We find that from low to high redshift, the faint end slope becomes steeper, $M^{*}$ remains roughly constant, and $\phi^{*}$ is declining.

\item The SMFs of galaxies in high and low densities are very similar, with the exception of massive galaxies living in denser environments at redshifts of $z<1.5$.

\item The impact of environmental processes on the shape of the galaxy SMFs of quenched galaxies is tested and we find tentative evidence for $\alpha$ being steeper in higher densities at $0.5<z<1.5$. The value of $M^{*}$ is higher for red systems in high density which suggests high density is causing more efficient growth of massive, quenched galaxies.

\item Results from the literature are confirmed regarding the double Schechter function nature of both the total and quenched SMFs. In addition, we find that the form of the galaxy SMF of galaxies with $n>2.5$ is best described as a double Schechter function at $z<2$. This suggests mass dependant evolution of these galaxies. 

\item The double Schechter function form of the $n>2.5$ SMF is due to a combination of a quenched (red) high stellar mass population, and a star forming (blue) low mass population. This suggests that links between growth of a bulge and the cessation of star formation are only present at the high-mass end.

\end{itemize}
If we are to understand the growth of galaxy stellar mass then the form of the galaxy SMF with respect to different properties needs to be understood. To advance in this field larger data sets are needed to improve the number statistics and reduce the uncertainties when computing number densities in sub-populations. Furthermore, higher resolution images are necessary to look at galaxy structure and future telescopes such as the James Webb Space Telescope (JWST), Euclid, and the European Extremely Large Telescope (E-ELT) will provide the quality of data required to advance these studies.

\section*{Acknowledgments}
A.M. acknowledges funding from the STFC and a European Research Council Consolidator Grant (P.I. R. McLure). We also acknowledge funding from the Leverhulme trust.

\bibliographystyle{mnras}
\bibliography{refs}

\label{lastpage}
\end{document}